\documentclass[iop]{emulateapj}
\pdfoutput=1
\usepackage{amsmath,amssymb}
\usepackage{color}
\usepackage{natbib}
\usepackage{graphicx}
\usepackage{hyperref}
\usepackage{ulem}
\usepackage{adjustbox}
\usepackage[draft]{todonotes}
\usepackage{mathtools}

\shorttitle{Metacalibration} \shortauthors{Huff and Mandelbaum}

\begin{document}
\title{Metacalibration: Direct Self-Calibration of Biases in Shear Measurement}
\author{Eric M. Huff\altaffilmark{1}}
\author{Rachel Mandelbaum\altaffilmark{2}}

\altaffiltext{1}{Jet Propulsion Laboratory, California Institute of Technology,
4800 Oak Grove Drive, Pasadena, CA 91109, USA}
\altaffiltext{2}{McWilliams Center for Cosmology, Department of Physics, Carnegie Mellon University,
  Pittsburgh, PA 15213, USA}

\keywords{cosmology: observations --- gravitational lensing: weak ---
  methods: observational}

\begin{abstract}
  One of the primary limiting sources of systematic uncertainty in forthcoming weak lensing
  measurements is systematic uncertainty in the quantitative relationship between the distortions
  due to gravitational lensing and the measurable properties of galaxy images. We present a
  statistically principled, general solution to this problem. Our technique infers multiplicative
  shear calibration parameters by modifying the actual survey data to simulate the effects of a known
  shear. It can be applied to any shear estimation method based on weighted averages of galaxy shape
  measurements, which includes all methods used to date for shear estimation with real data.  Use of
  the real images mitigates uncertainty due to unknown galaxy morphology, which is a serious concern
  for calibration of shear estimates based on image simulations.  We test our results on simulated
  images from the GREAT3 challenge, and show that the method eliminates calibration biases for
  several different shape measurement techniques at the level of precision measurable with the
  GREAT3 simulations (a few tenths of a percent).
\end{abstract}

\section{Introduction}

Accurate measurement of weak gravitational lensing offers the most
direct probe of the dark sector of the universe
\citep[e.g.,][]{2001PhR...340..291B,2003ARA&A..41..645R,schneider06,2008ARNPS..58...99H,2010RPPh...73h6901M,2013PhR...530...87W}.
Weak lensing measurements are thus a core part of the international
cosmology program, and a key science driver for several wide-field
astronomical imaging cameras and associated surveys -- the Kilo Degree
Survey\footnote{\url{http://kids.strw.leidenuniv.nl}}
\citep{KiDS_main}, the Dark Energy Survey \citep{DES_main}, the
Hyper Suprime-Cam and associated survey\footnote{\url{http://hsc.mtk.nao.ac.jp/ssp/}}
\citep{HSC_main}, LSST\footnote{\url{http://www.lsst.org/lsst/}}
\citep{2009arXiv0912.0201L},
Euclid\footnote{\url{http://sci.esa.int/euclid/}\url{http://www.euclid-ec.org}}
\citep{2011arXiv1110.3193L}, and
WFIRST\footnote{\url{http://wfirst.gsfc.nasa.gov}}
\citep{2015arXiv150303757S}.

Despite this investment, the weak lensing community has more work to
do in order to ensure that the algorithms for inferring shear are
unbiased at the required levels to avoid systematic errors from
dominating over the statistical errors. One of the largest such
systematic error sources is the {\it shear calibration bias}, the
quantitative relationship between the true gravitational lensing shear 
and its observables as estimated from the ensemble of galaxies in the survey.

In the weak shear limit that is most relevant for wide-field cosmology, the
gravitational lensing signal can be described as a linear
transformation
$\boldsymbol{A}\mathbf{x}_{\rm true} = \mathbf{x}_{\rm obs}$ between
the lensed and unlensed image coordinates, parameterized by two shears
$(\gamma_1,\gamma_2)$ and a convergence $\kappa$
\begin{align}
\boldsymbol{A}=
\begin{pmatrix}
1 + \kappa + \gamma_1 & \gamma_2 \\
\gamma_2 & 1 + \kappa - \gamma_1
\end{pmatrix}
\end{align}.
The major observable effect of weak lensing is to perturb the measured
ellipticities $\mathbf{e} = (e_1,e_2)$ of galaxies. At large
separations, these shapes no know preferred direction, so the mean
$\mathbf{e}$ should vanish over a wide enough field. Weak lensing
studies exploit this intrinsic symmetry, and search for spatially
coherent anisotropies in the ensemble of observed galaxy shapes
arising from lensing distortions produced by foreground matter.

The effects of the shear and convergence on this observable cannot be
straightforwardly distinguished, so the fundamental quantity
constrained by lensing is the reduced shear
\begin{align}
 g = \frac{\gamma}{1 - \kappa}.
\end{align}

The responses of individual galaxy images to $g$ vary depending on the
choice of ellipticity measure and the intrinsic shape and orientation
of each galaxy. Lensing studies rely on ensemble averages of galaxy
ellipticities, and the shears are weak enough that these ensemble
averages usually respond linearly to an applied (reduced) shear, so it
is conventional to define the multiplicative shear calibration and
additive bias parameters as
\begin{align}
\langle \mathbf{e}\rangle = (1 + m)\langle \mathbf{g}\rangle + \mathbf{c}
\label{eqn:calibPars}
\end{align}
where $\mathbf{e}$ and $\mathbf{g}$ are ensemble-averaged shears and
ellipticity measures, respectively. Generally, $\mathbf{c}$ is a
result of measurement biases (such as an incomplete correction for the
point-spread function) that introduce a preferred direction in the
image plane. It can in principle be known or removed with sufficient
knowledge of the experiment. $m$ depends in part on the ensemble of
(unobserved) galaxy properties, so it is impossible in principle to
know exactly {\it a priori} (though \citealt{2014MNRAS.438.1880B} show
how to derive this information for their proposed shear estimator,
which does not make use of an ensemble average over ellipticities,
from deeper calibration fields).

In practice, a nonlinear response generically introduced by the algorithms
used for measurement of $\boldsymbol{e}$ can introduce both
multiplicative and additive biases in a manner that interacts with the
unknown true ensemble properties of galaxies
\citep{2007MNRAS.380..229M,2011MNRAS.414.1047Z}, and are very
difficult to predict from first principles. For this reason, the weak
lensing community has organized a series of blind measurement
challenges, where participants attempted to extract an unkown lensing
signal from simulated images.  The earliest of these were the first
two Shear TEsting Programmes \citep[STEP1,
STEP2]{2006MNRAS.368.1323H,2007MNRAS.376...13M}. The results made two
things clear: that lensing measurement algorithms needed to improve in order to avoid being
systematics-dominated,
and that shear measurement was sufficiently complex that successive
simulation challenges should focus on a subset of the issues.

The next round of simulation challenges -- GREAT08, GREAT10, and GREAT3
\citep{2009AnApS...3....6B,2013ApJS..205...12K, 2015MNRAS.450.2963M} --
embraced a narrower focus and saw significant performance
improvements. They also drove improvements in our understanding of 
various sources of bias in shear estimation, which is of use in future
algorithmic development.  The best-performing algorithms from the most
recent challenge, GREAT3, reduced $m$ and $c$ to levels approaching
those needed for the most ambitious planned lensing measurements,
albeit with simulations that did not include all the features of real
data.

While this was certainly good news, the narrowed focus of the GREAT
challenges necessarily left some of the most important sources of
lensing calibration bias untouched. Remaining issues of significant
concern include biases resulting from:
\begin{itemize}
\item object detection and selection
\item deblending
\item wavelength-dependent effects
\item instrumental defects and nonlinearities
\item star-galaxy separation
\item non-white pixel noise
\item cosmic rays and other image artifacts
\item redshift-dependent calibration biases
\item shear estimation for low-resolution and/or low signal-to-noise ratio ($<12$) galaxies
\end{itemize}
The impact of these factors depends strongly on the specifics of the
experiment. For this reason, shear calibration in current and future
experiments relies heavily on simulations designed to match the
properties of each experiment
\citep{KiDS450,2016MNRAS.tmp..827J}. Such external simulations are
always limited in their realism:  accurately modeling everything
relevant about the experiment turns out to be extremely difficult. Showing that a given simulation suite
is adequate for calibrating a lensing measurement is a formidable
challenge in its own right (c.f.\ the Ultra Fast Image Generation
simulations described in \citealt{2013A&C.....1...23B}, or the
calibration simulations used for the KiDS weak lensing cosmology in
\citealt{2016arXiv160605337F}).

The method outlined in this paper is motivated by the observation that
introducing a synthetic shear signal into real data is much easier
than building a realistic comprehensive first-principles simulation
suite. While in practice the need for accurate simulations of the
ensemble of galaxy images is sometimes met by relying on images from
external deep fields like the Hubble Space Telescope's COSMOS survey
\citep{2007ApJS..172..196K,2007ApJS..172....1S,2007ApJS..172...38S},
the deeper fields needed for calibration of future surveys like LSST
and WFIRST may not be available in the volume necessary.

 Perturbing the actual data automatically incorporates features
present in real images (e.g., image artifacts, selection biases,
unusual high-redshift galaxy morphologies) that are otherwise
difficult to accurately simulate.  It enables the determination of how the {\em real galaxy
  population} in the data responds to a shear directly and empirically.


We have implemented this concept, which we call metacalibration, using
the public GalSim \citep{2015A&C....10..121R} image simulation
package, and designed our algorithm to wrap an arbitrary external
shear estimation module, provided that it functions by estimation of per-galaxy ellipticities and
then estimates the ensemble shear through weighted averages. We test our technique on simulated GREAT3
image data, and find that it successfully calibrates several older shear
estimation methods to a level of accuracy comparable to the
best-performing algorithms from the GREAT3 challenge. We also
demonstrate that our algorithm can detrend additive biases resulting
from incomplete point-spread function (PSF) corrections by introducing synthetic PSF
ellipticity.  We make our metacalibration scripts available for
general use.

\section{Method}
There are three layers to the shear calibration method we propose
here. The first is the generation of the modified images using a procedure similar to one proposed
in \cite{2000ApJ...537..555K}.
We use the GalSim package \citep{2015A&C....10..121R} to modify real
astronomical images by adding synthetic shear and PSF distortions of
known amplitude. These modified images are counterfactuals; they are a
model for what would have been observed under (nearly) the same image
quality conditions, on the same galaxies, with a different shear. If
the measurement process is repeated on the counterfactual images, the
result gives an accurate estimate of the response of the galaxy population to a shear.

The second layer is the choice of ellipticity measure used to estimate
per-object shapes. This step is the primary focus of most studies that
address shear calibration biases. Here we are agnostic about the
choice of measurement algorithm; as long as the algorithm is
sufficiently well-behaved (in a manner that we will describe in
Sec.~\ref{subsec:shapemeas}), the image manipulation step can be used
to generate an accurate shear responsivity.

The final layer is the choice of averaging mechanism to estimate the
response of the ensemble shear estimate to an applied shear. Noise
properties of shape measurements can vary widely depending on the
shape measurement method, which entails similar variation in the
metacalibration estimates for shear responsivity. For the cases we
describe below, an optimal strategy for ensemble averaging produces
significant gains over more straightforward averaging schemes.

\subsection{Generating a Counterfactual Image}
\label{sec:counterfactual}
Fortunately, for the weak shears under consideration in most
cosmological survey applications, the relationship between the shear
and the galaxy shapes (or related observables) is very close to
linear, so accurate shear calibration requires only the first
derivative of the galaxy properties with respect to the shear. What
follows is a method for estimating this derivative directly from the
images. Throughout we will assume that the observed image
$I({\mathbf{x}})$ is equal to the unsmeared galaxy image
$G(\mathbf{x})$ convolved with some point-spread function (including the
atmospheric seeing, the optical PSF, and the pixel response function) $P(\mathbf{x})$.

In an ideal world, we would calibrate our measurement algorithm by
making measurements while varying the gravitational shear experienced
by the pre-seeing image, constructing the counterfactual image
$I'(\mathbf{x}| {\boldsymbol g})$:
\begin{equation}
  I'({\mathbf{x}}|\mathbf{g}) = P \ast\left( \hat{\mathbf{s}}_{\mathbf{g}}G\right)
\end{equation}
where $\hat{\mathbf{s}}_{\boldsymbol g}$ is the shear operator that
produces the reduced shear $\mathbf{g}$, as in
e.g.~\cite{2002AJ....123..583B}. The shear sensitivity of the image
would then be a straightforward numerical derivative of $I'$ with
respect to $\mathbf{g}$, and the shear sensitivity of an ellipticity
measure $\mathbf{e}$ can be calculated from measurements on multiple
counterfactual images. We can even write down a formal procedure for
producing $I'$ from $I$ if we know $P$:
\begin{equation}
  I'({\mathbf{x}}|\mathbf{g}) = P \ast \left[\hat{\mathbf{s}}_\mathbf{g}\left( P^{-1} \ast I \right)\right].
\end{equation}
The convolutions become products in Fourier space, where we can write
\begin{align}
\tilde{I'({\mathbf{k}}|\mathbf{g})} = \tilde{P}^{\ast} (\mathbf{k}) \: \hat{\mathbf{s}}_\mathbf{g}\left(\frac{\tilde{I} (\mathbf{k})}{\tilde{P}^\ast (\mathbf{k})}\right)
\end{align}
Noise in the original image $\tilde{I}$ generally has power at Fourier
modes where $\tilde{P}$ is small or vanishing. The power in these
modes will thus be formally large or infinite. Because of the shear
operation, this power is not subsequently cancelled by multiplication
by $\tilde{P}$. We must choose a new PSF $\Gamma$ for the final
convolution step to suppress this deconvolution-amplified noise.

If $||\tilde{P}(\mathbf{k})||$ is monotonically decreasing with $k$,
this condition can be achieved without introducing additional PSF
anisotropy by choosing
\begin{align}
\Gamma(\mathbf{x}) = P\left((1+2|\gamma|)\mathbf{x}\right).
\end{align}
This does not always work, however. If $||\tilde{P}(\mathbf{k})||$
crosses zero (as in cases with a strongly under-sampled PSF) the
ratio of $\tilde{\Gamma}(\mathbf{k})$ and the sheared, deconvolved
image will still be formally large or infinite, as power from
$k-$values beyond the zero crossing will be dragged by the shear
operation into the region where the dilated PSF does not vanish.

Other, implementation-specific considerations may be important when
choosing $\Gamma$. When choosing a target PSF, it may prove convenient
to design one which is well-suited to the shear estimator in hand. We
defer exploration of this topic to future work.

Our chosen procedure for producing a sheared counterfactual image is
\begin{equation}
I'({\mathbf{x}}|\mathbf{g}) = \Gamma \ast \left[\hat{\mathbf{s}}_\mathbf{g} \left(P^{-1} \ast I \right)\right].
\end{equation}
This procedure clearly requires a good model for $P$, but so do all
shear measurements. PSF model errors enter at the
same order in measurements on the resulting image that they would in
an unmodified image.

Once the counterfactual image $I'(\mathbf{x}|\mathbf{g})$ with
$\|\mathbf{g}\| \ll 1$ has been created, the galaxy detection and
shear measurement pipeline should be rerun. This provides a measure of
the shear sensitivity -- not for the original image, but for an image
with the PSF $\Gamma$. This requires that the full measurement -- not
just the sensitivity analysis -- be run on an additional
counterfactual image $I'(\mathbf{x}|\mathbf{g}=0)$, so that the numerical
derivative $\frac{\partial I'}{\partial \mathbf{g}}$ is well-defined.

This procedure introduces anisotropic correlated noise, which can
produce a systematic multiplicative shear bias. If the noise
properties of the initial image are known, the noise anistropy can be
removed with the addition of further anisotropic correlated noise
(with power spectrum carefully chosen). As we describe below, we have
not found noise isotropization to be a necessary step for the images
that we used for testing.  These have an effective $S/N$ limit of
$\sim 12$, and the mode of the distribution is $\sim20$. Concurrent
work \citep{metacalII} investigates the effects of the anisotropic
correlated noise at lower signal-to-noise ratios, and describe
effective mitigation procedures.

Metacalibration can be used to mitigate other systematics as
well. Even those measurement methods with the highest scores in the
GREAT3 lensing challenge were unable to completely remove the effects
of PSF ellipticity on the inferred shear. We can introduce an
artifical PSF anisotropy by replace $\Gamma$ with a PSF containing the
desired synthetic distortion.  We show below that reconstructing
images with added PSF ellipticity, rather than added shear, allows us
to de-trend some of the bias due to PSF anisotropy. A similar approach
could be used to measure additive or multiplicative calibration biases
arising from any effect -- signal or systematic error -- that can be
simulated by perturbing the images as above.

\subsection{Shape Measurement Algorithms}
\label{subsec:shapemeas}
Accurate ensemble shears can only be derived through measurement of
the counterfactual images described above if the shape measurement
algorithm is sufficiently well-behaved. Here, that entails the
requirement that the quantity reported by the shape measurement
algorithm be sufficiently linear in the underlying shear in the regime
relevant for the measurement that the ensemble response is truly
linear.


We test a variety of shape algorithms below that make use of differing
definitions of ellipticity. As we are attempting to construct a shear
calibration procedure that is agnostic about the choice of per-object
shape measurement algorithm, and which only requires that we use a
measured galaxy property with approximately linear sensitivity to
shear (called a shape measure), we will use $\mathbf{e}$ below to
signify all of the shape measures discussed in this paper, regardless
of their precise definition.

\subsection{Ensemble Shear Inference}
Counterfactual images can in principle be used to derive a per-object
shear response for a modified version of the original image.  However,
the quantities of interest in Eq.~\ref{eqn:calibPars} are ensemble
responses.  Hence it is necessary to run the shear inference
step on the counterfactual images, in order to get the shear
responsivity of the ensemble shear estimate.

The distribution of measured shear responses determines the nature of
the ensemble inference procedure. The distribution of MetaCal
responses, especially for shape measures that involve ratios of noisy
quantities, can make simple averaging schemes problematic; power-law
tails result in a very high variance and slow convergence of the mean
response. We develop and implement a simple technique below which
deals adequately with the large noise in the regaussianized moment
responses. If the measurement algorithm is nonlinear, however, then
our averaging scheme will require some additional correction, beyond
what we develop here, for the resulting nonlinear ensemble response.

\subsection{Algorithmic Limitations}
The fundamental assumptions of the image processing steps are
frequently violated in real data. The image manipulation step assumes
that the image is linearly related to the true surface brightness on
the sky. This is not valid for common image processing artifacts like
cosmic rays, for saturated pixels, or when charge-deflection effects
produce a flux-dependent PSF (e.g.,
\citealt{2015JInst..10C5032G}). 

The presumption of a single shear, with a single response factor, can
also be problematic. The lensing signal varies with redshift along a
single line of sight, so blended images of multiple galaxies at
appreciably different redshifts involve at least two different shears,
and the relationship of the metacalibrated response to the underlying
shear field is not straightforward. Nevertheless, these issues are
generic, and will if unaddressed will cause problems for any shear
inference method; we require our images to be reliable representations
of the sky, and that the shear field be in some sense single-valued
where it can be measured.

Linearity in the ensemble inference is also vital if the image
processing algorithm only computes first derivatives of the image with
respect to the shear, as in the implementation we describe here. More
finite difference steps could in princple be used to calibrate a
nonlinear shear response, but this significantly increases the noise
and the computation cost of the method. Here we will perform enough
finite difference steps to allow for an estimate of the linearity of
each estimator.

\section{Implementation}
\subsection{Image Modification}\label{subsec:imagemod}
We use
GalSim\footnote{\url{https://github.com/GalSim-developers/GalSim}} to
manipulate the images and to generate simulations for validation. For
each galaxy, we create nine modified images: two for each of the two
shear components, two for each of the two PSF ellipticity components,
and one for the final measurement using the enlarged PSF $\Gamma$
(two-sided derivatives with respect to shear and PSF ellipticity were
found to be less noisy than one-sided derivatives).  We run the
provided shape measurement pipeline on each of these images, and the
results are used to construct a set of finite difference estimates of
shear calibration and additive PSF biases.

This sort of image manipulation is trivial to carry out using GalSim;
we rely on the rigorous testing of the image convolution,
interpolation, and resampling algorithms that the development team
performed to enable the GREAT3 shear testing simulations.  From the
perspective of numerical validation, the tests in section~9 of
\cite{2015A&C....10..121R} illustrate that GalSim can accurately
render sheared images of quite complex galaxy and PSF light profiles
with its default interpolants and settings that control numerical accuracy.

For each galaxy and PSF postage stamp, we first create an
\texttt{InterpolatedImage} object. This object is deconvolved by the
PSF model (including the pixel response). For the shear finite
differences, we apply a small shear $\Delta\mathbf{g}$ (typically 1\%)
to the resulting deconvolved image. The original PSF is dilated by
twice the shear distortion to produce $\Gamma$, and then re-convolved
with the sheared deconvolved image. This reconvolved, sheared image is
then passed to the shape measurement routine, along with the image
representation of the new, enlarged PSF $\Gamma$. For the PSF
sensitivity, we follow a similar procedure, but shear the dilated PSF
image rather than the deconvolved galaxy image. Finally, we create a
reconvolved image with no added shear but with the PSF $\Gamma$, on
which we perform the final shape measurement.


\begin{figure*}
\begin{center}
\includegraphics[width=0.45\textwidth]{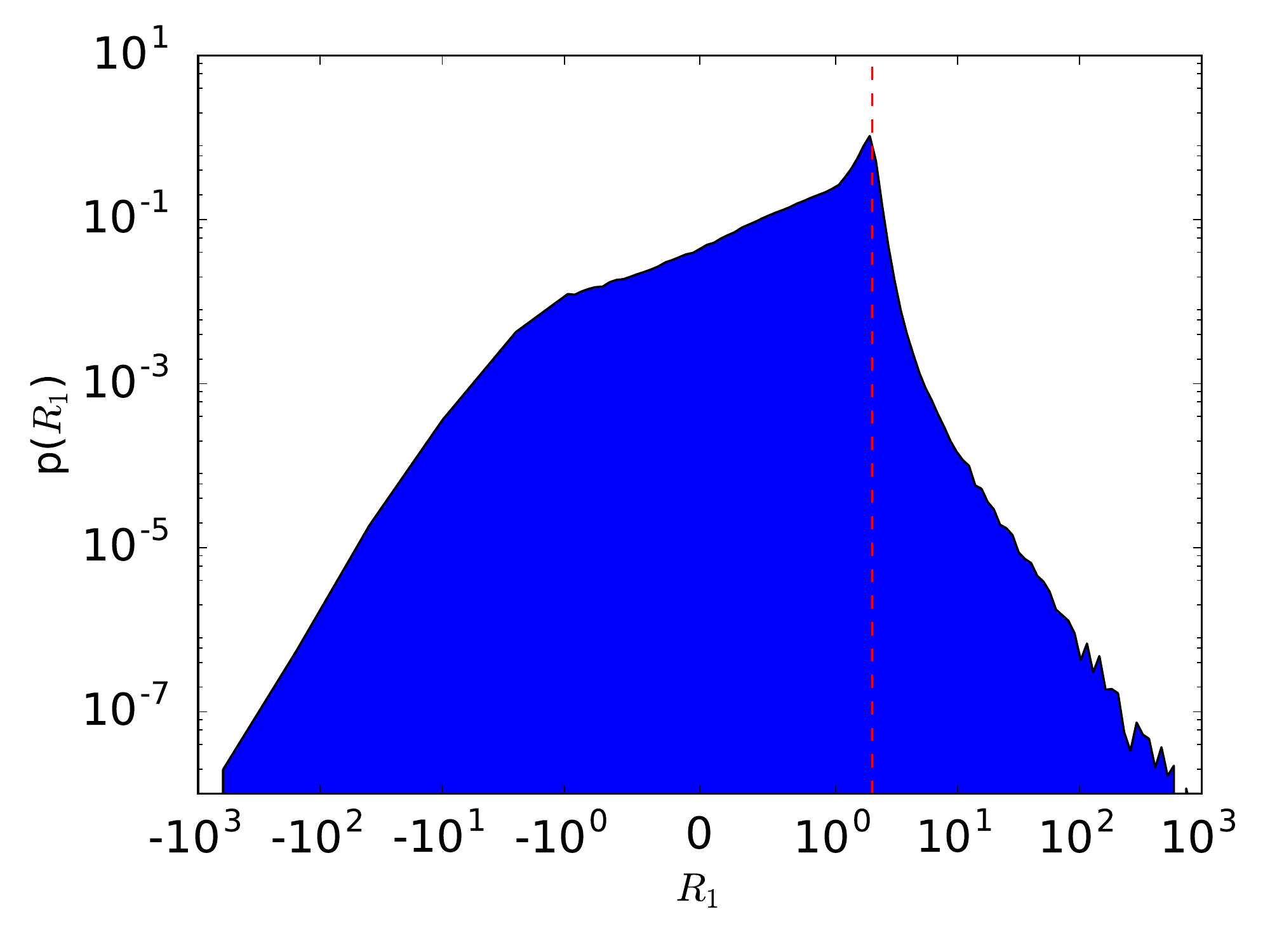}
\includegraphics[width=0.45\textwidth]{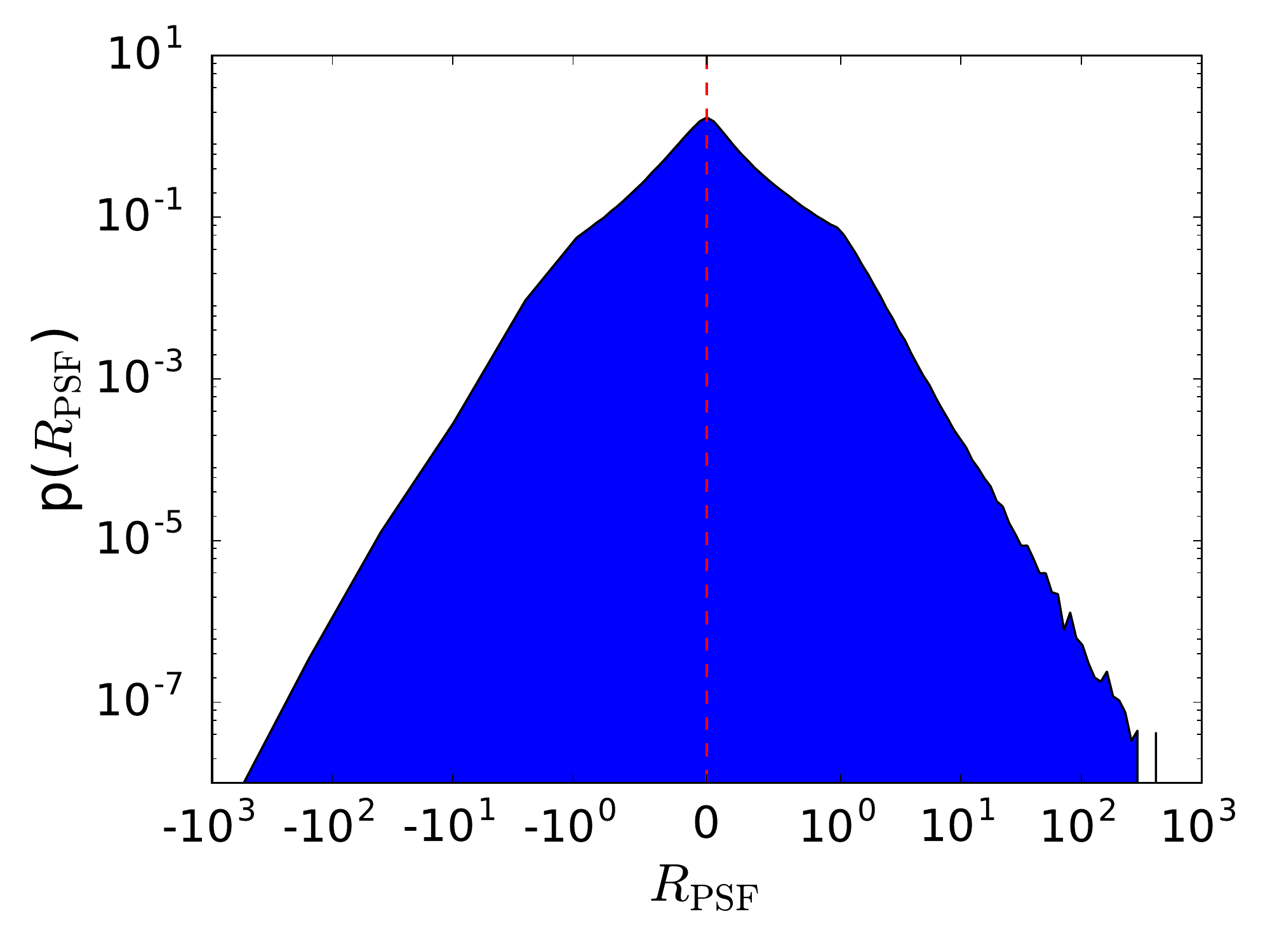}
\end{center}
\caption{{\bf Left:} Normalized distribution of metacalibration shear
  responsivities from regaussianization, on the
  Control-Ground-Constant branch of the GREAT3 simulations.  {\bf
    Right:} Distribution of metacalibration PSF ellipticity
  responsivities from regaussianization, on the
  Control-Ground-Constant branch of the GREAT3 simulations. A
  vertical red dashed line is drawn for reference at the expected
  responsivity for perfectly round objects, $R=2$, in the left
  panel. }
\label{fig:calibhist}
\end{figure*}

\subsection{Shear Estimation Algorithms}

Since the metacalibration method can in principle be used to calibrate
shears from any shear estimation algorithm derived using an average of per-object shapes, we chose three easily
available shear estimation methods, all of which are implemented in
GalSim.  Two of these methods are traditional shear estimation
methods that have somewhat different assumptions but are both based on
object moments.  One method is not a standard shear estimation method
at all: we use linear combinations of the directly observed second
moments without any correction for the PSF.  In principle, the
information about how those respond to shear should be determined by
metacalibration to correctly infer the shear.  The difference in this
case is that instead of providing a small correction to the outputs of
a PSF correction method, we rely on metacalibration to do the entirety
of the PSF correction, which is a very stringent test that may at least 
partially violate some of the assumptions about the ensemble average of the measured quantities having a linear
response to shear.  The three methods are described below.

\subsubsection{Regaussianization}

Re-Gaussianization \citep{2003MNRAS.343..459H} is a PSF correction
method based on the use of the moments of the image and of the PSF to
correct for the effects of the PSF on the galaxy shapes. It includes
corrections for the non-Gaussianity of the galaxy profile
\citep{2002AJ....123..583B,2003MNRAS.343..459H} and of the PSF (to
first order in the PSF non-Gaussianity). The performance of this
algorithm has been extensively studied in real data and simulations
\citep[e.g.,][]{2005MNRAS.361.1287M,2012MNRAS.420.1518M,2013MNRAS.432.1544M,2015MNRAS.450.2963M}. 

The outputs of the re-Gaussianization algorithm are PSF-corrected
``distortions'', which for an object with purely elliptical isophotes
with minor-to-major axis ratio $q$ and position angle $\theta$ with
respect to the $x$ axis in pixel coordinates are defined as
\begin{equation}
(e_1, e_2) = \frac{1-q^2}{1+q^2}\left(\cos{2\theta},\sin{2\theta}\right).
\end{equation}
As discussed in \cite{2002AJ....123..583B}, the response of a
distribution of galaxies with some intrinsic distribution of
distortions $p(e)$ to a shear 
depends on
the $p(e)$ itself.  Conceptually, we can think of an ensemble shear
estimator using re-Gaussianization outputs as
\begin{equation}
\hat{g}_j = \frac{\langle e_j\rangle}{\mathrm{d}\langle e_j\rangle/\mathrm{d}g_j}
\end{equation}
where the denominator gives the response of the ensemble average
distortion to a shear (often called the responsivity).  Estimators of
this shear responsivity use the observed galaxy $p(e)$ and its
moments, and for typical $p(e)$, the denominator is around
$1.7$--$1.8\approx 2 (1-e_\text{RMS}^2)$ in terms of the per-component
RMS distortion. As this implementation was meant to be a simple and
fast example, its intrinsic calibration correction is a simple one
that does not include all known systematics.

\subsubsection{KSB}

The KSB method \citep{1995ApJ...449..460K} parametrises galaxies and
stars according to their weighted quadrupole moments.  The main
assumption of the KSB method is that the PSF can be described as a
small but highly anisotropic distortion convolved with a large
circularly symmetric function.  With that assumption, the shear can be
recovered to first-order from the observed ellipticity of each galaxy
via
\begin{equation} \label{eqn:weight}
g=P_{g}^{-1}\left(e^{\rm obs}-\frac{P^{\rm sm}}{P^{\rm sm*}}e^{*}\right),
\end{equation}
where asterisks indicate quantities that should be measured from the
PSF model at that galaxy position, $P^{\rm sm}$ is the smear
polarisability (see \citealt{2006MNRAS.368.1323H} for definitions) and
$P_g$ is the correction to the shear polarisability that includes
the smearing with the isotropic component of the PSF. The
ellipticities are constructed from weighted quadrupole moments, and
the other quantities involve higher order moments. A circular Gaussian
weight of scale length $r_g$ is used, where $r_g$ is galaxy size, as
determined by the second moment of the surface-brightness profile. 

The KSB method returns a per-object estimate of the shears
$(\hat{g}_1, \hat{g}_2)$. We can use metacalibration to
remove multiplicative and additive biases that come from averaging the
per-object KSB shear estimates.

\subsubsection{Linear Moments}

As mentioned previously, the third method we use does not involve
PSF-corrected galaxy shapes.  Instead, we use linear combinations of
the second moments of galaxy images.  The motivation behind this
choice is as follows.  One way to estimate the distortion $(e_1,e_2)$
is via combinations of the second moments of the light profile,
\begin{equation}
\langle x_i\rangle = \frac{\int x_i w({\mathbf x}) I({\mathbf x}) \mathrm{d}^2{\mathbf x}}{\int w({\mathbf x}) I({\mathbf x}) \mathrm{d}^2{\mathbf x}}
\end{equation}
for $i=1, 2$,
\begin{equation}
M_{ij} = \frac{\int (x_i-\langle x_i\rangle)(x_j-\langle x_j\rangle) w({\mathbf x}) I({\mathbf x}) \mathrm{d}^2{\mathbf x}}{\int w({\mathbf x}) I({\mathbf x}) \mathrm{d}^2{\mathbf x}}
\end{equation}
for $i,j=1,2$, and finally 
\begin{equation}\label{eq:moments-div}
e_1 = \frac{M_{11}-M_{22}}{M_{11}+M_{22}}, \qquad e_2 =\frac{2M_{12}}{M_{11}+M_{22}}.
\end{equation}

One source of noise (and noise bias) in traditional moments-based methods is the
division of two noisy quantities in Eq.~\ref{eq:moments-div},
typically followed by further division by other noisy quantities to
remove the dilution of the galaxy shape by the PSF.  Thus, as a final
example of a statistic that we will attempt to use as a calibrated
shear estimator with metacalibration, we define the following linear
combinations of moments:
\begin{equation}
\hat{M}_i = (M_{11}-M_{22}, 2M_{12}).
\end{equation}

Clearly these moments are sensitive to a number of nuisance
quantities, like the galaxy flux and size, and the PSF size and shape.
In principle, metacalibration should be able to nonetheless determine
the response of this statistic to shear,
$\mathrm{d}\hat{M}_i/\mathrm{d}g$, and produce a reliable shear
estimate from the ensemble-averaged $\hat{M}_i$ values, provided that
the linear model for the signal and dominant sources of systematic
error is correct.  This is a quite stringent test of the
metacalibration method, as it is unclear whether that purely linear
model will be valid in this case.

\subsection{Per-Object Responsivity}
Shape measurements on the set of modified images can be used to derive
noisy shear and PSF responsivities for individual galaxies. In the
case where the measured ellipticity is thought to depend linearly on
the lensing shear and the PSF ellipticity, the shapes measured from
the sheared, reconvolved images with positive and negative applied
shears, $\mathbf{e}_+$ and $\mathbf{e}_-$, can be used for a
straightforward finite-difference estimate of the per-object shear
response
\begin{align}
R &= \frac{\partial \mathbf{e}}{\partial \mathbf{g}}  \\
 &=\frac{\mathbf{e}_{+} - \mathbf{e}_{-}}{2\Delta\mathbf{g}}
\end{align}
Additive biases introduced by the shape measurement can be extracted from the
sum of these two quantities
\begin{align}
\mathbf{c} = \frac{\mathbf{e}_+ + \mathbf{e}_-}{2} - \mathbf{e},
\end{align}
where the final $\mathbf{e}$ on the right-hand side is the ellipticity
measured using the reconvolved PSF $\Gamma$, not that measured from
the unmodified original image.

If the shape measurement algorithm does not perfectly remove PSF
ellipticity, then the shapes measured from shearing the PSF
($\mathbf{e}_{+,\rm PSF}$ and $\mathbf{e}_{-,\rm PSF}$) permit
calculation of at least the linear-order residual PSF ellipticity
biases:
\begin{align}
R_{\rm PSF} = \frac{\mathbf{e}_{+,\rm PSF} - \mathbf{e}_{-,\rm PSF}}{2\Delta\mathbf{g}}.
\end{align}
The result of this process is a catalog of per-object shear and PSF
responsivities every galaxy.  A histogram of these quantities for one
of the measurement algorithms examined in this paper is shown in
figure~\ref{fig:calibhist}. 

Accurate calibration depends on characterising the ensemble response,
however, not the per-object responses. We argue that the power-law
wings of the distribution make estimation of the ensemble response by
simple averaging problematic, and describe next a general scheme for
regularizing the ensemble response estimation. These wings are
a consequence of the fact that the regaussianization shapes are
themselves ratios of noisy quantities; other methods with more compact
support may not require this sort of regularization scheme.

\subsection{Ensemble Responsivity}
We test the metacalibration procedure on two different shear
estimation methods -- regaussianization and KSB -- each of which is
known to have calibration biases that depend on the $S/N$, galaxy size
and morphology, and so on. For both methods, we use the entire
ensemble of measured shapes to build a model of the {\it unlensed}
shape distribution, $p_0(\mathbf{e})$. There is no guarantee that the
average shear (or PSF ellipticity) over the ensemble is actually small
enough that the effective mean ellipticity is zero, however. Before
proceeding with the inference, we subtract from each galaxy the
expected contribution from the PSF for its field,
$R_{\rm PSF}e_{\rm PSF}$. We then symmetrize the resulting
distribution by averaging it with its reflection about $\mathbf{e}=0$
(note that the resulting model for the null shape distribution will be
somewhat broader than the true distribution of ellipticities at zero
shear). The newly symmetrized model unlensed distribution is
$p_{0,\rm sym}$. We perform the same PSF subtraction from the measured
catalog prior to performing inference in the procedure described
below.


If the measured shape distribution $n(\mathbf{e}_{\rm meas})$ is linear
in the shear, then we can write
\begin{align}
  \frac{n(\mathbf{e}_{\rm meas})}{N_{tot}} = p_{0,\rm sym}(\mathbf{e}) + \mathbf{g}\cdot \partial_{g} p_{0,\rm sym}(\mathbf{e})
\end{align}
It will be convenient to discretize this distribution into a
histogram. If the probability of a galaxy ending up in the
$i^{\rm th}$ shape histogram bin is $q_i$, and each galaxy's shape can
be considered an independent draw from some underlying distribution,
then the likelihood function for an observed histogram is the
multinomial likelihood
\begin{align}
p( \left\{N_i\right\} |\left\{ q_i\right\} ) = \frac{N_{\rm tot}!}{\prod\limits_i (N_i!)}\prod\limits_j q_j^{N_j}
\label{eqn:multnomial}
\end{align}
where $N_{\rm tot} = \sum\limits_i N_i$ is the total number of samples
in the histogram. The covariance matrix for the
bin amplitudes of this histogram is
\begin{align}
{\rm cov}(N_i, N_j) = C_{ij} = \begin{cases}
  q_i(1-q_i)N_{\rm tot}, & i=j \\
  -q_iq_j N_{\rm tot}, & i \neq j.
\end{cases}
\end{align}
To make the notation for what follows less cumbersome, let the
normalized histogram be $h_i = N_i / N_{\rm tot}$, and its first
derivative with respect to the shear be
$\mathbf{\Delta}_h=\partial_{g}\mathbf{h}_{\rm fid}$.

Given a measured shape histogram with some unknown shear and a
fiducial, unlensed shape histogram, the (component-wise)
minimum-variance estimator for $g$ is
\begin{align}
\hat{g} = \frac{\mathbf{\Delta}_h^T C^{-1}\left( \mathbf{h}_{\rm meas} - \mathbf{h}_{\rm fid}\right)} {\mathbf{\Delta}_h^TC^{-1}\mathbf{\Delta}_h},
\label{eqn:hist_est}
\end{align}
with variance 
\begin{align}
\sigma^2_{\hat{g}} = \frac{1}{\mathbf{\Delta}_h^TC^{-1}\mathbf{\Delta}_h}
\label{eqn:hist_est_var}
\end{align}

This method for shear inference has as its tunable parameter only the
histogram binning scheme. Once we have chosen a suitable scheme, we
then bin the prior into equal-number bins and calculate its shear
derivative 
by adding a small shear $\mathbf{g}$ using the per-object responses,
then rebinning. This provides the inputs for the per-field shear
estimation.  To carry out the per-field shear estimation, we calculate
a shape histogram with these bins for each separate field, and
evaluate equations~\ref{eqn:hist_est} and~\ref{eqn:hist_est_var} using
the globally-derived $\mathbf{\Delta}_h^T$.

\subsection{Validating the Shear Inference}
\label{sec:model_checking}


If we have a poor model for the unlensed histogram, $h_{\rm fid}$,
then the results for shear inference will be biased. Once we have derived a mean shear for
each field, we can quantify the degree to which the shape histogram
resembles our model of the unlensed prior distribution.

To do this, we subtract from each galaxy in each field the product of
its individual responsivity and the estimated shear (the PSF
contribution having already been removed, as above). We then define
the distance between each field and the unlensed prior using
equation~\ref{eqn:multnomial}, taking the probabilities $q_i$ from the
unlensed model and the histogram amplitudes from the current field,
after correcting for the estimated shear. If the shear response
measured for the unlensed prior is correct, then the performance of
the estimator will depend only on the similarity of the model to the
measurement field. The likelihood can then be used as an objective
criterion for the quality of the inference.

In practice, we only apply a cut on the likelihood in cases where we have an {\it a
  priori} reason to believe that the prior constructed as we describe
above will not be an adequate description of all of the simulation
fields. In our tests below, we apply this cut in simulation branches
where there is large expected variation in the PSF properties. In
these cases, we exclude from our analysis those fields with the $10\%$
worst log-likelihood cuts.



We can also devise a simple test for the linearity (in the shear) of
the ensemble estimator. The estimator we use to infer the shear
response relies on two-sided finite-difference numerical derivatives,
but it is also possible to get a noisier measure the nonlinear
ensemble response by testing for consistency between the forward- and
backwards finite-difference estimates for the ensemble response. We
quantify this with the parameter $\eta_{\rm nl}$, defined as
\begin{align}
1+\eta_{nl} = \frac{\Delta_{h+}^T C^{-1} \Delta_{h-}}{\Delta_{h}^TC^{-1}\Delta_{h}}
\end{align}
where the $+$ and $-$ subscripts refer to the forward- and backwards
finite difference histogram derivative estimates.

\subsection{Relationship to previous implementations}

As shown in the GREAT3 results paper \citep{2015MNRAS.450.2963M}, an
early version of metacalibration was used in the GREAT3 challenge.
That implementation differs from the one presented here and released
publicly in association with this paper in two important ways.  First,
the model for systematics was simpler than the one presented below; it
neglected additive systematics entirely, focusing exclusively on
multiplicative systematics. Second, the method for inferring shears
from an ensemble of objects was entirely different.  These differences
are of sufficient importance that the GREAT3 results (especially the
ones for additive systematics) are not relevant to the implementation
described here.

\section{Testing Framework}
We test the performance of our algorithm on simulated image sets. Our
baseline simulations are drawn from the GREAT3 simulation suite. We
run additional simulations in order to distinguish between potential
biases arising separately from the three steps in our inference
framework.

\subsection{Simulated Images}

We use the GREAT3 simulation framework as the source of simulated
images that we use for testing purposes.  For more detail about that
simulation framework, see the GREAT3 handbook
\citep{2014ApJS..212....5M} and results paper
\citep{2015MNRAS.450.2963M}, or the publicly available
software\footnote{\url{https://github.com/barnabytprowe/great3-public}}.

In brief, we use simulated ``branches'' containing 200 ``subfields''.
Each subfield contains $10^4$ galaxies placed on a $100\times 100$
grid; the galaxies in a given subfield all have the same (unknown)
shear and the same known PSF.  The galaxy population within a subfield
follows a distribution of signal-to-noise ratio, size, ellipticity,
and morphology based on that in the {\it Hubble Space Telescope} ({\it
  HST}) COSMOS survey
\citep{2007ApJS..172..196K,2007ApJS..172....1S,2007ApJS..172...38S},
roughly approximating a galaxy sample with a depth of $I<25$.  To
ensure that most methods will be able to measure all galaxies, an
effective signal-to-noise cut\footnote{This was initially advertised as a cut at 20, however the
  GREAT3 results paper shows that for a more realistic signal-to-noise estimator, the effective cut
  is around 12.} of $\gtrsim 12$ and a minimal resolution
cut was imposed (resulting in different sets of galaxies in subfields
that have different-sized PSFs).  90$^\circ$ rotated pairs of galaxies
were included, to cancel out shape noise \citep{2007MNRAS.376...13M}.
The PSF in the simulations comes from the combination of an optics
model from a ground-based telescope, along with a Kolmogorov PSF with
a typical ellipticity variance.  Thus, the galaxy and PSF properties
are non-trivially complicated.  The noise is stationary Gaussian
noise.  The ultimate goal is to estimate the average shear in each
subfield in an unbiased way, without any multiplicative bias or
correlations with the per-subfield PSF ellipticity.


We consider two sets of galaxy populations.  One comes directly from
{\it HST} images, and includes a process to remove the {\it HST} PSF before
shearing (both operations being carried out in Fourier space) and
convolving with the final target PSF \citep{2012MNRAS.420.1518M}.  The
other galaxy population consists of simple parametric representations
of those {\it HST} images.  These populations have the same effective
distributions of size, ellipticity, and so on, but one includes
realistic galaxy morphology while the other only includes such realism
as can be captured by the sum of two S\'{e}rsic profiles.  In the
language of GREAT3, we use simulations corresponding to
control-ground-constant (describing the parametric galaxy sample,
ground-based simulated data, with a constant per-subfield shear) and
real$\_$galaxy-ground-constant (the realistic galaxy sample), denoted
CGC and RGC.

\subsection{Simulation Branches}
\label{sec:branches}

The simplest of our simulation tests was performed on a
newly-generated set of simulations that is closely analogous to the
GREAT3 CGC branch (parametric galaxy profiles), but with one
modification to avoid a problem raised in the results paper
\citep{2015MNRAS.450.2963M}.  There, it was noted that the CGC branch
has a small fraction of outlier fields related to unusually large optical PSF
aberrations, specifically defocus and trefoil.  Thus, our first
simulated dataset is designed exactly like CGC but with all
aberrations in the optical PSF set to zero, to ensure reasonable
consistency of data quality.  Note that the atmospheric PSF is still
drawn from a distribution of seeing values for each subfield. For this
branch, it is not necessary to use a likelihood cut to remove fields
with aberrant PSF behavior in defining the null ellipticity
distribution, and so we include all branches in our analysis. In the
accompanying figures and table, this branch is denoted by
``CGC-noaber-regauss.''

The next branch we analyze is similar to the previous one, but with
realistic galaxies. This lets us separate the effects of realistic
galaxy morphology from the problems inherent in correcting a complex
PSF. Just as with the previous branch, is not necessary to use a
likelihood cut to remove fields with aberrant PSF behavior in defining
the null ellipticity distribution, and so we include all the generated
fields in our analysis. The results from this branch are labelled
``RGC-noaber-regauss.''

We use the GREAT3 CGC simulation branch as a baseline, and report the
performance of metacalibration on this branch for all three of our
chosen shape measurement algorithms; our results here are denoted in
the accompanying figures and tables with the labels ``CGC-regauss'',
``CGC-moments'', and ``CGC-ksb'', as appropriate.  Tests on this branch allow
straightforward comparison between the performance of our chosen
procedure and the variety of other algorithms tested in the GREAT3
challenge.

We next report results for the GREAT3 RGC branch (``RGC-regauss''),
incorporating more realistic galaxy morphologies along with the
aberrant PSFs introduced in the CGC branch. 

\begin{figure*}[t]
\begin{center}
\includegraphics[width=0.46\linewidth]{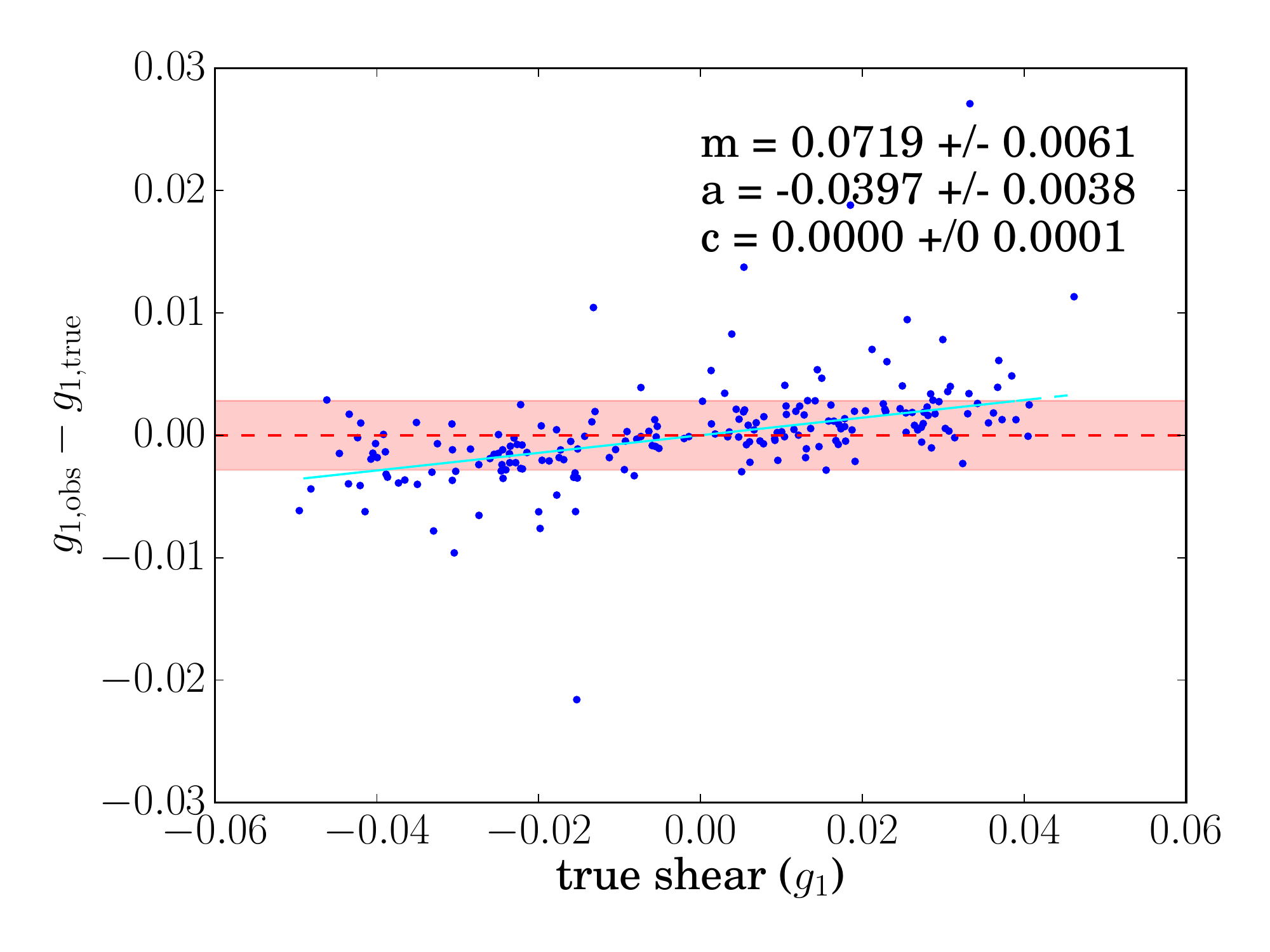}
\includegraphics[width=0.46\linewidth]{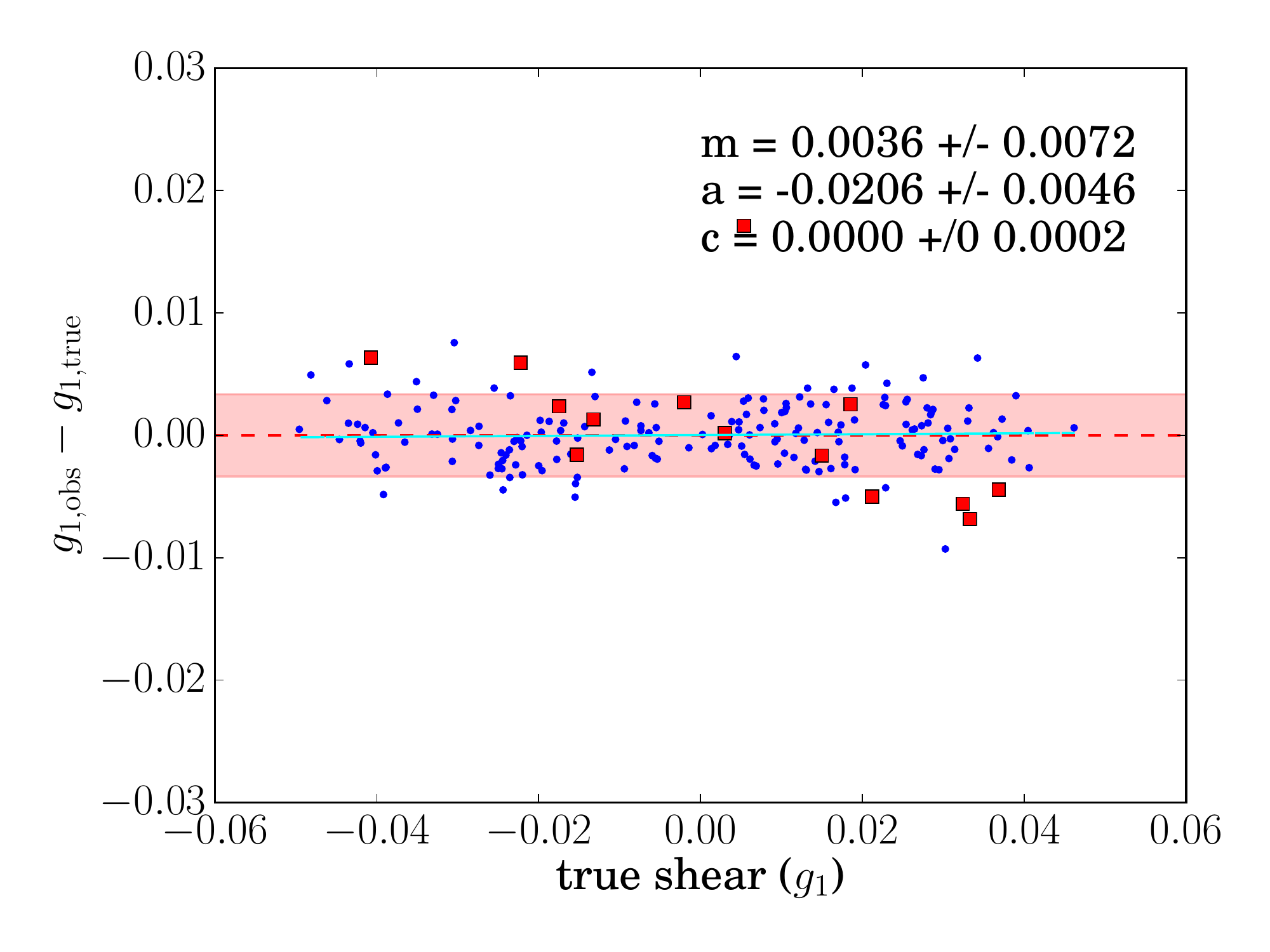}\\
\includegraphics[width=0.46\linewidth]{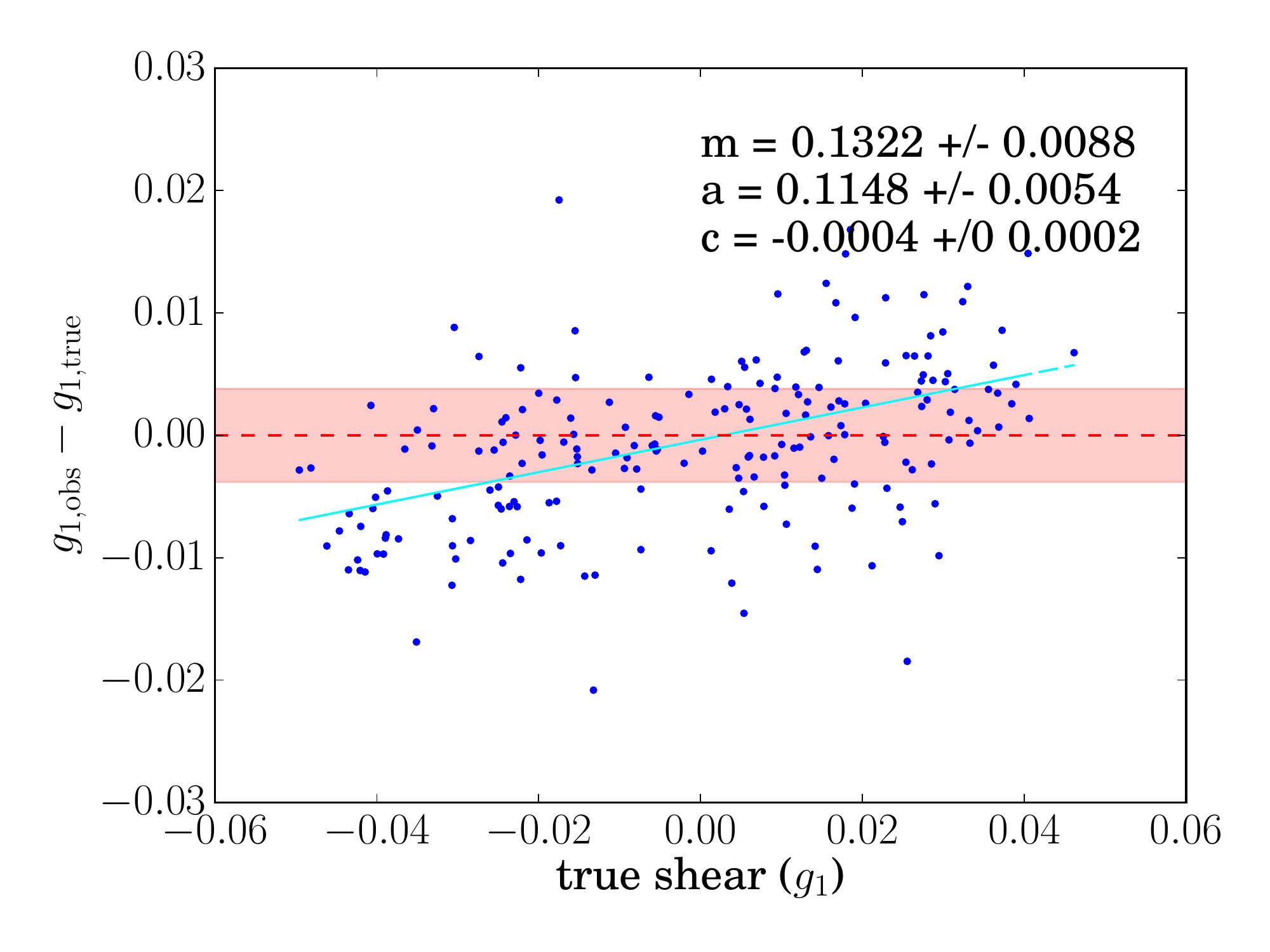}
\includegraphics[width=0.46\linewidth]{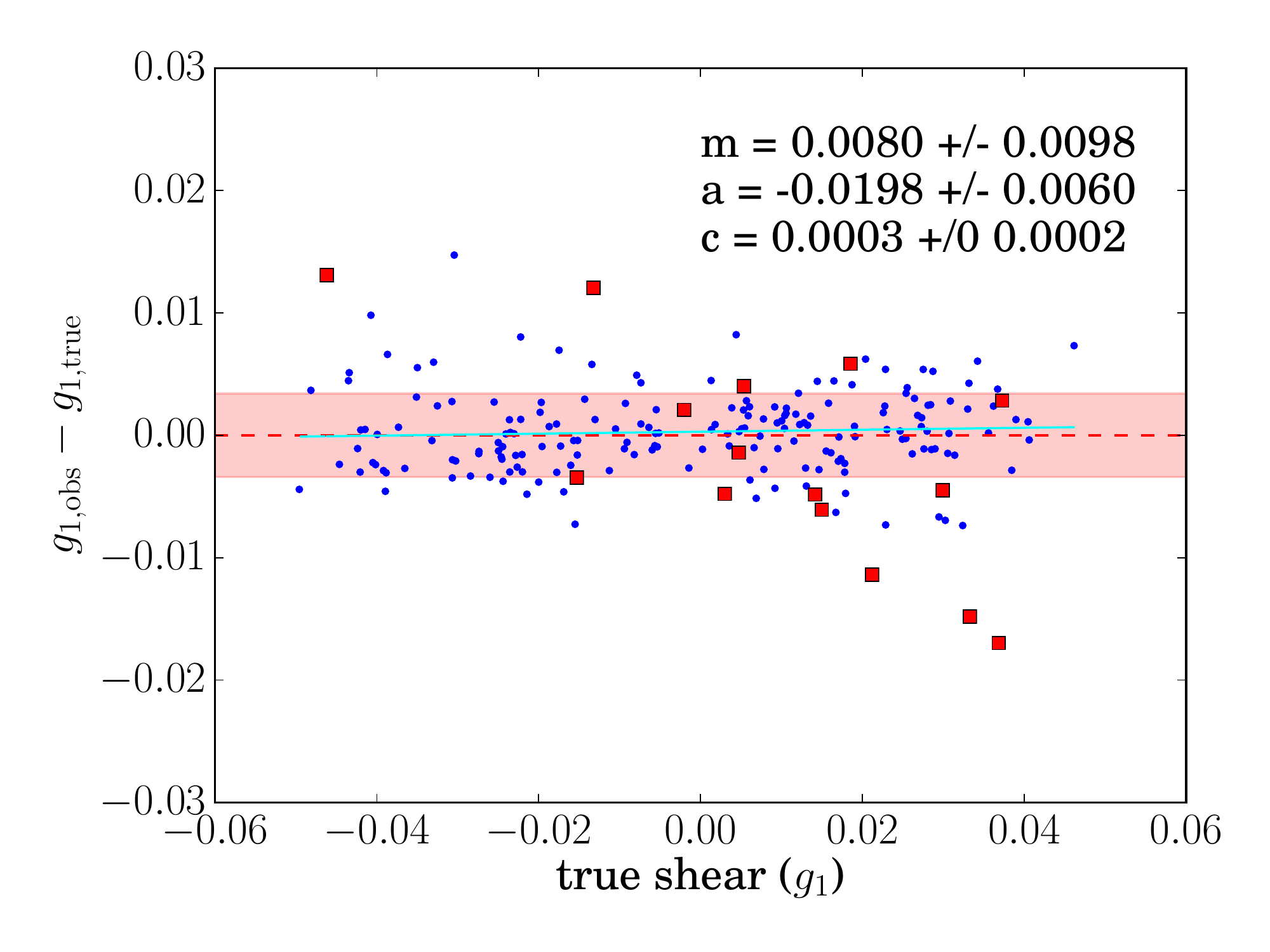}\\
\includegraphics[width=0.46\linewidth]{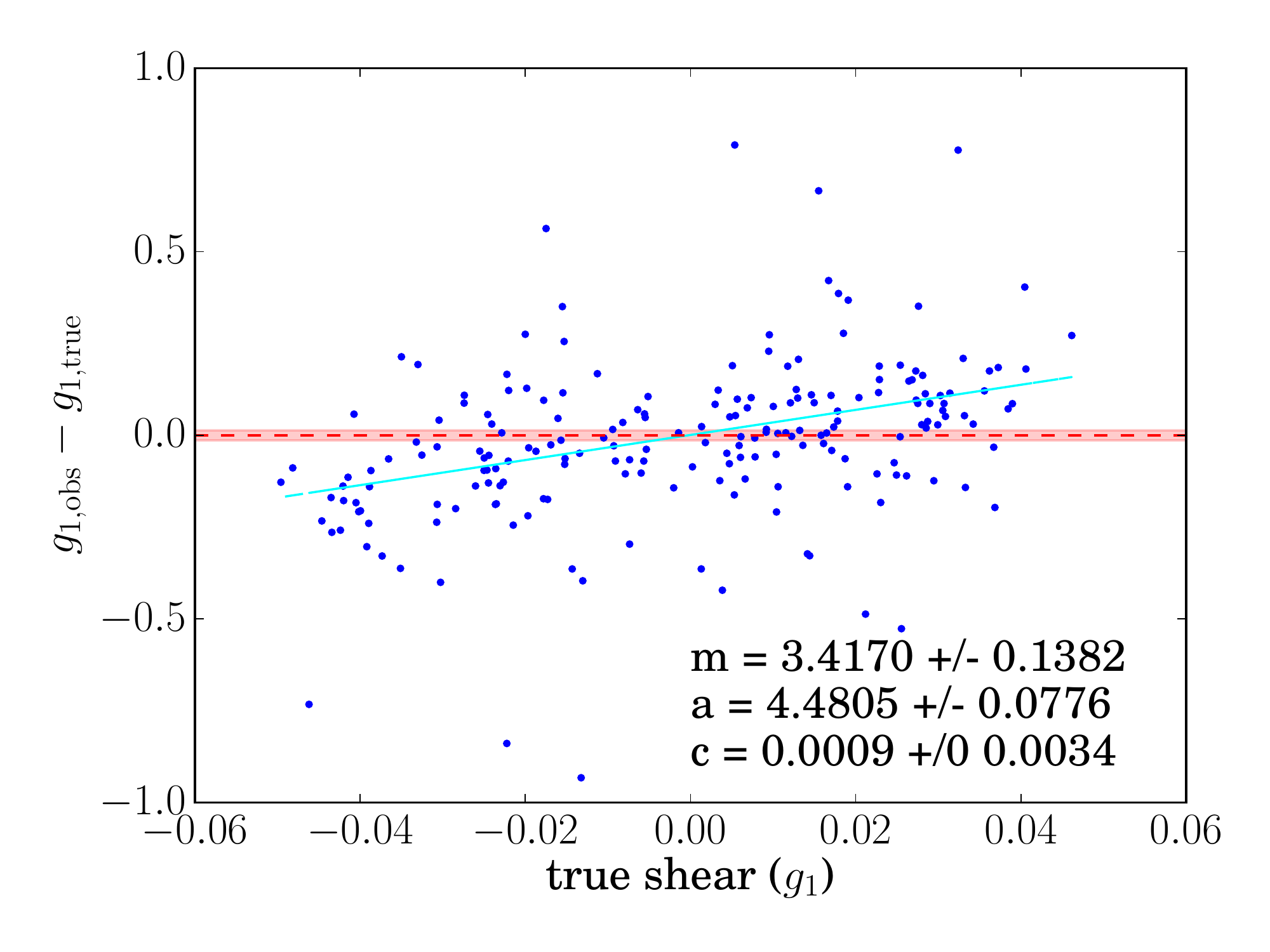}
\includegraphics[width=0.46\linewidth]{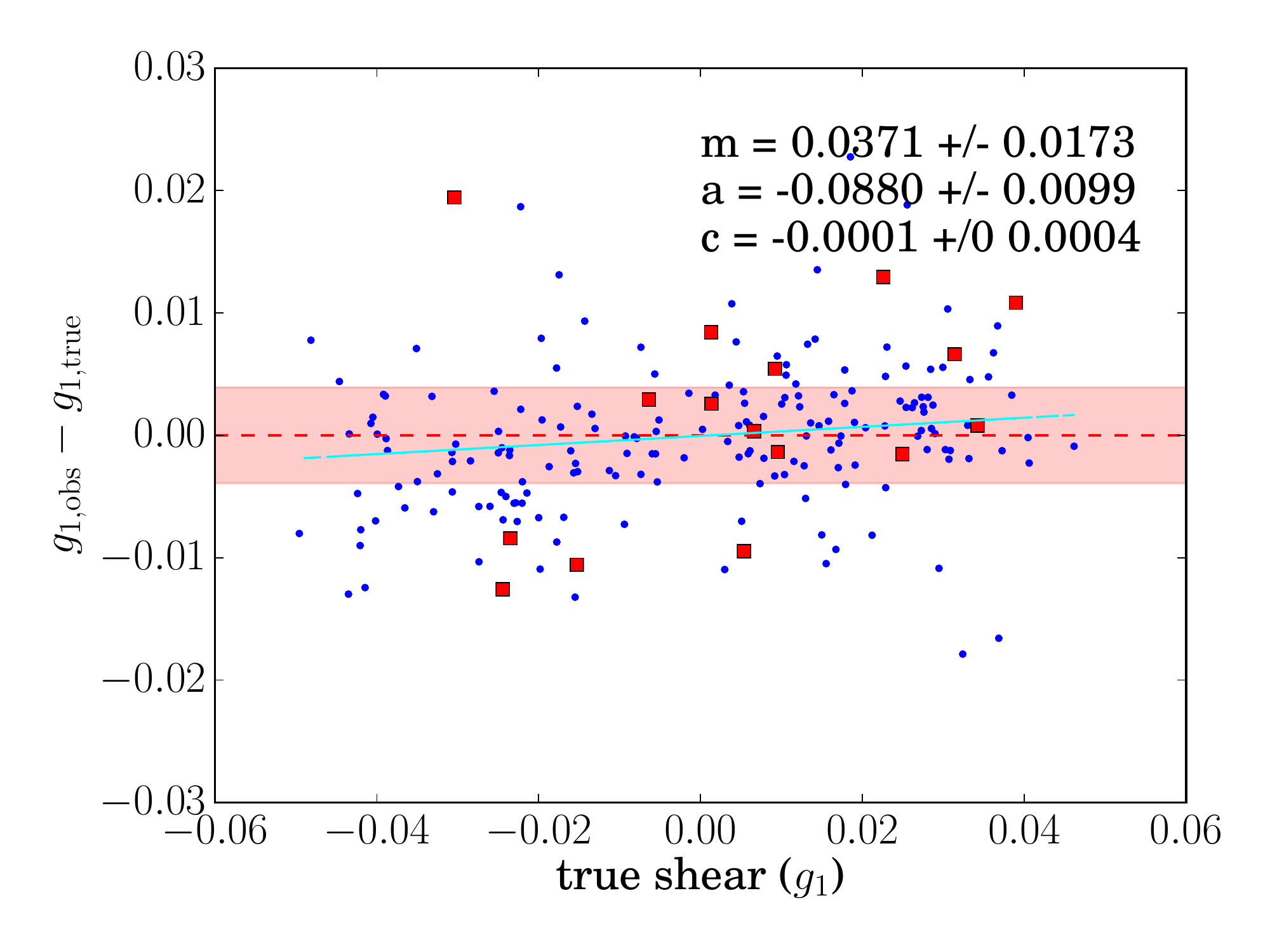}
\end{center}
\caption{Shear calibration bias $m_1$ {\it before} ({\bf left}) and
  {\it after} {\bf right} metacalibration for the regaussianization
  ({\bf top}), KSB ({\bf middle}), and Linear Moments ({\bf bottom})
  algorithms on the CGC branch. The shaded region covers the same
  vertical range in each panel.  Points excluded by the log-likelihood
  cut are marked with red squares.}
\label{fig:m_comparison}
\end{figure*}

One issue of concern is how to understand outlier fields.  In GREAT3,
there was a concern that some outliers were due to failure of our
model for interpreting the per-object shapes in fields that had large
aberrations. 
As a way to understand this, we generated a version of RGC that had
quite large aberrations that were identical in each field:
specifically defocus of $0.5$ waves and one component of trefoil of
$0.1$ wave.  An example of a PSF in one subfield is shown in
Fig.~\ref{fig:trefoil}; they do not all look identical, since the
atmospheric component was still allowed to vary stochastically
according to a distribution of seeing FWHM. This removes the difficulties
in building a null ellipticity distribution, isolating the impact of a
complex PSF. The results from this simulation are labelled
``RGC-FixedAber-regauss''.

Finally, we investigate the effects of increasing the noise in the
images. We re-run the ``CGC-regauss'' analysis five times, increasing
the noise in the initial image each time. The correlated noise
produced by the image modification procedure may affect the derived
calibration, and we expect this number of realizations to demonstrate
whether correlated noise biases is consistent with the expected 
signal-to-noise scaling. 
%

\begin{figure*}
\begin{center}
\includegraphics[width=0.4\linewidth]{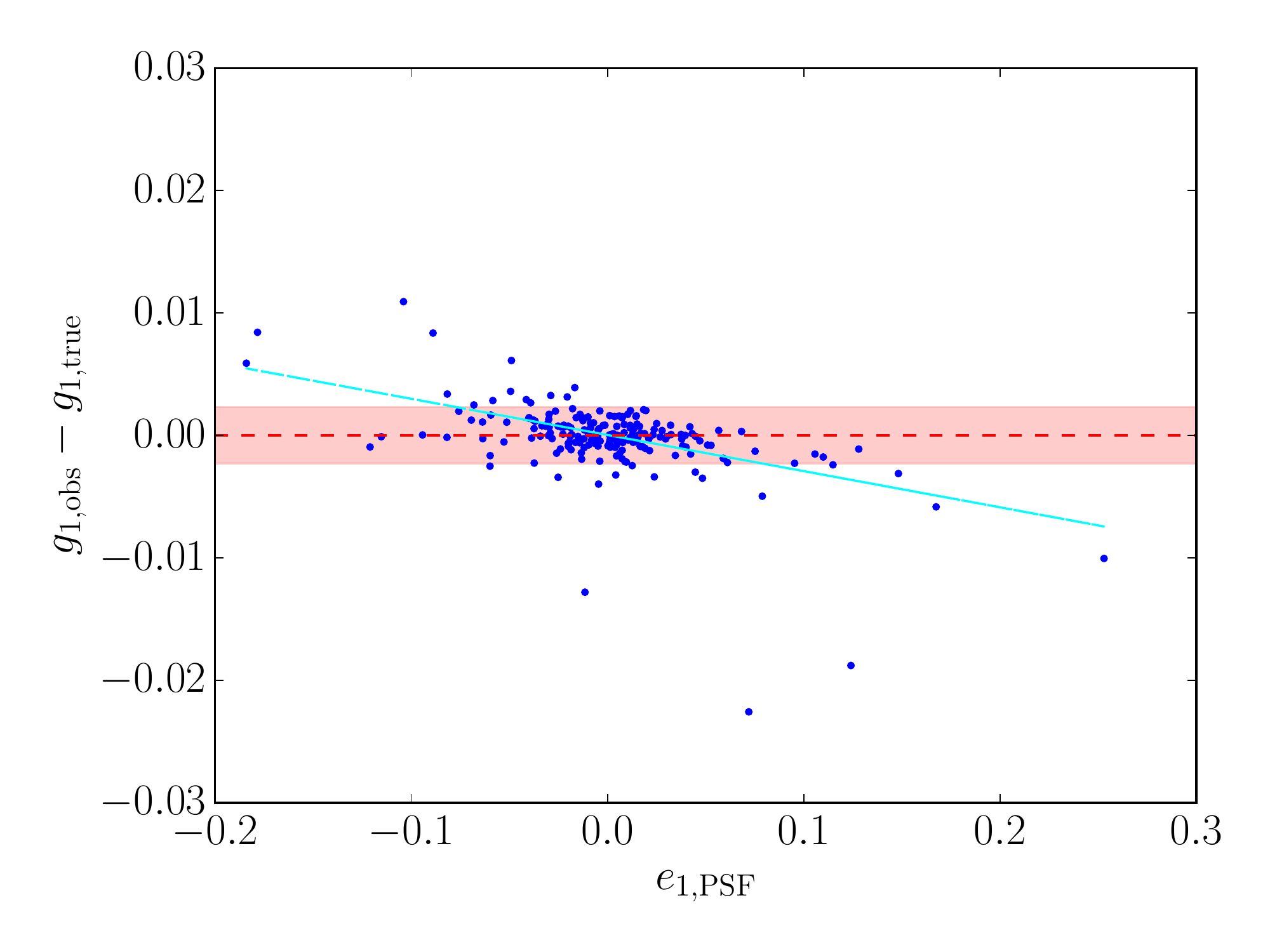}
\includegraphics[width=0.4\linewidth]{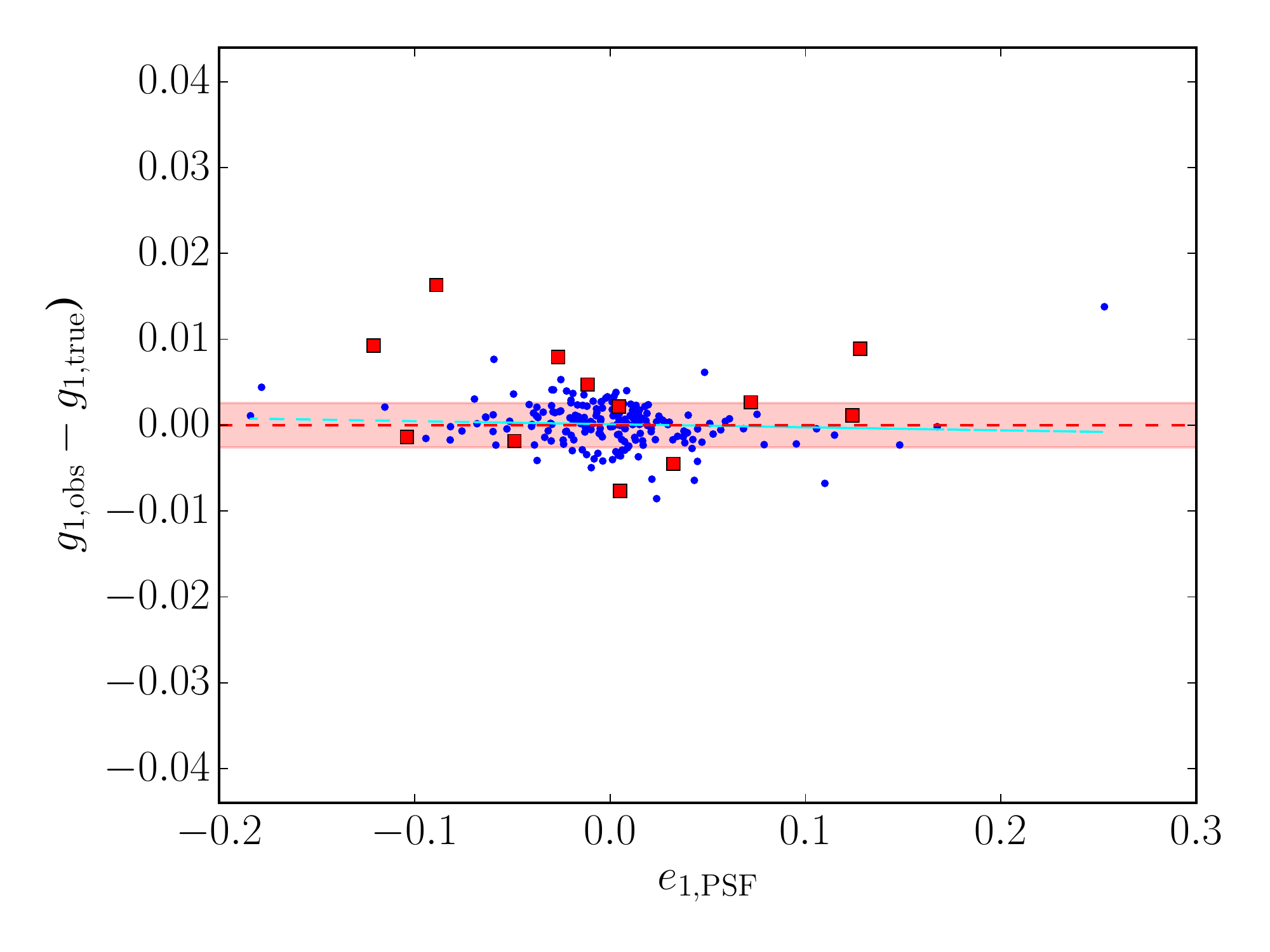}
\includegraphics[width=0.4\linewidth]{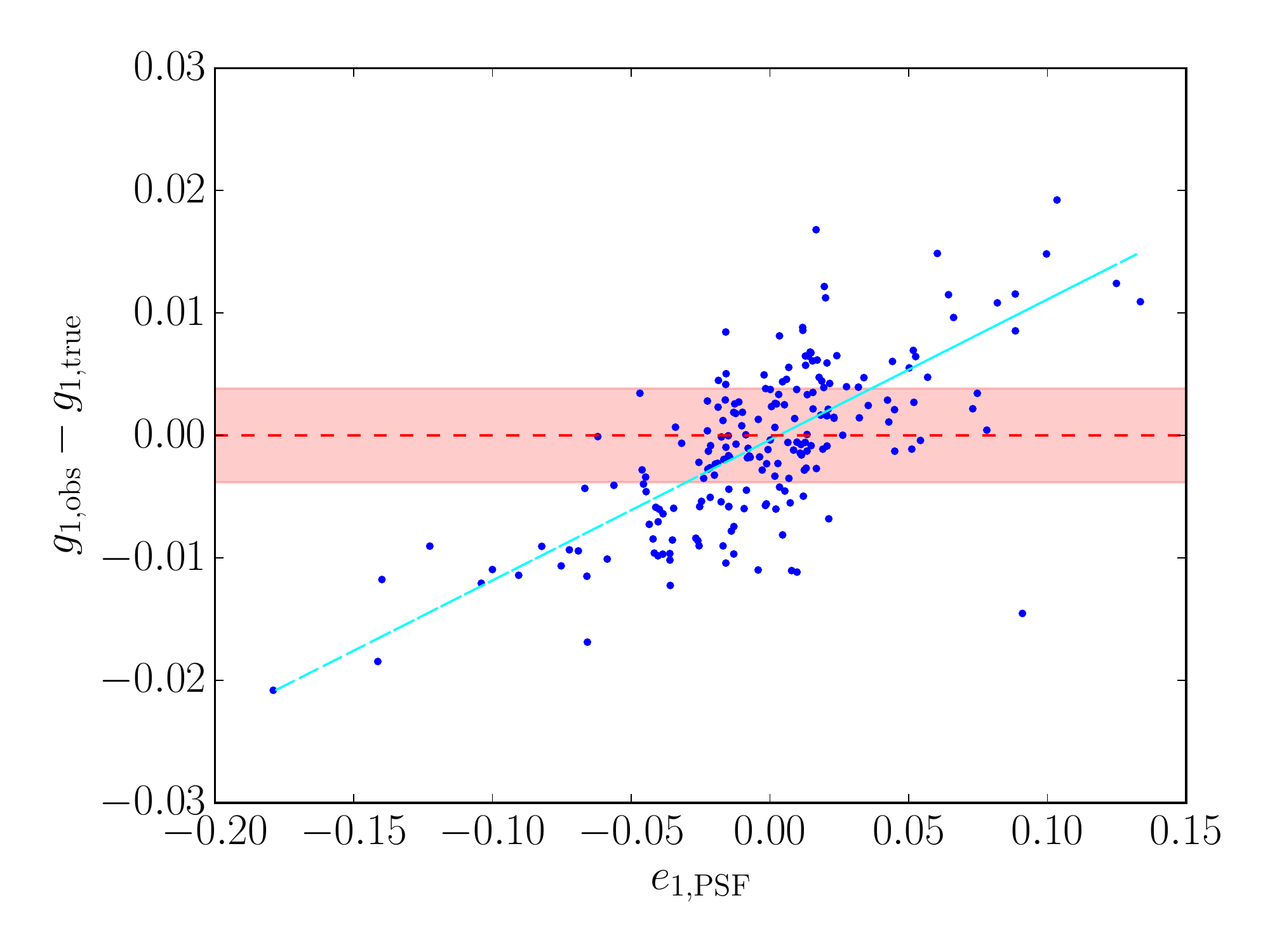}
\includegraphics[width=0.4\linewidth]{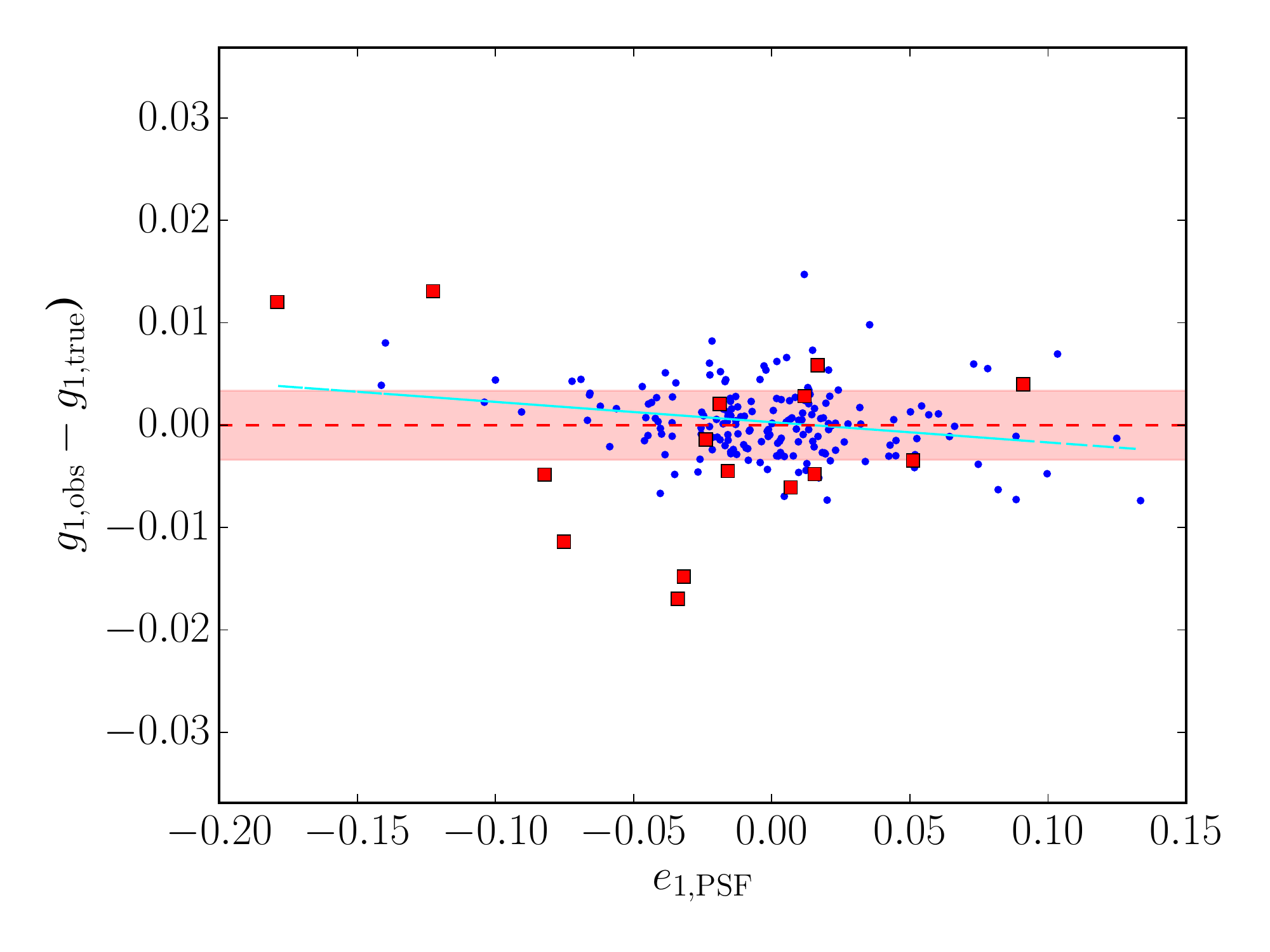}
\includegraphics[width=0.4\linewidth]{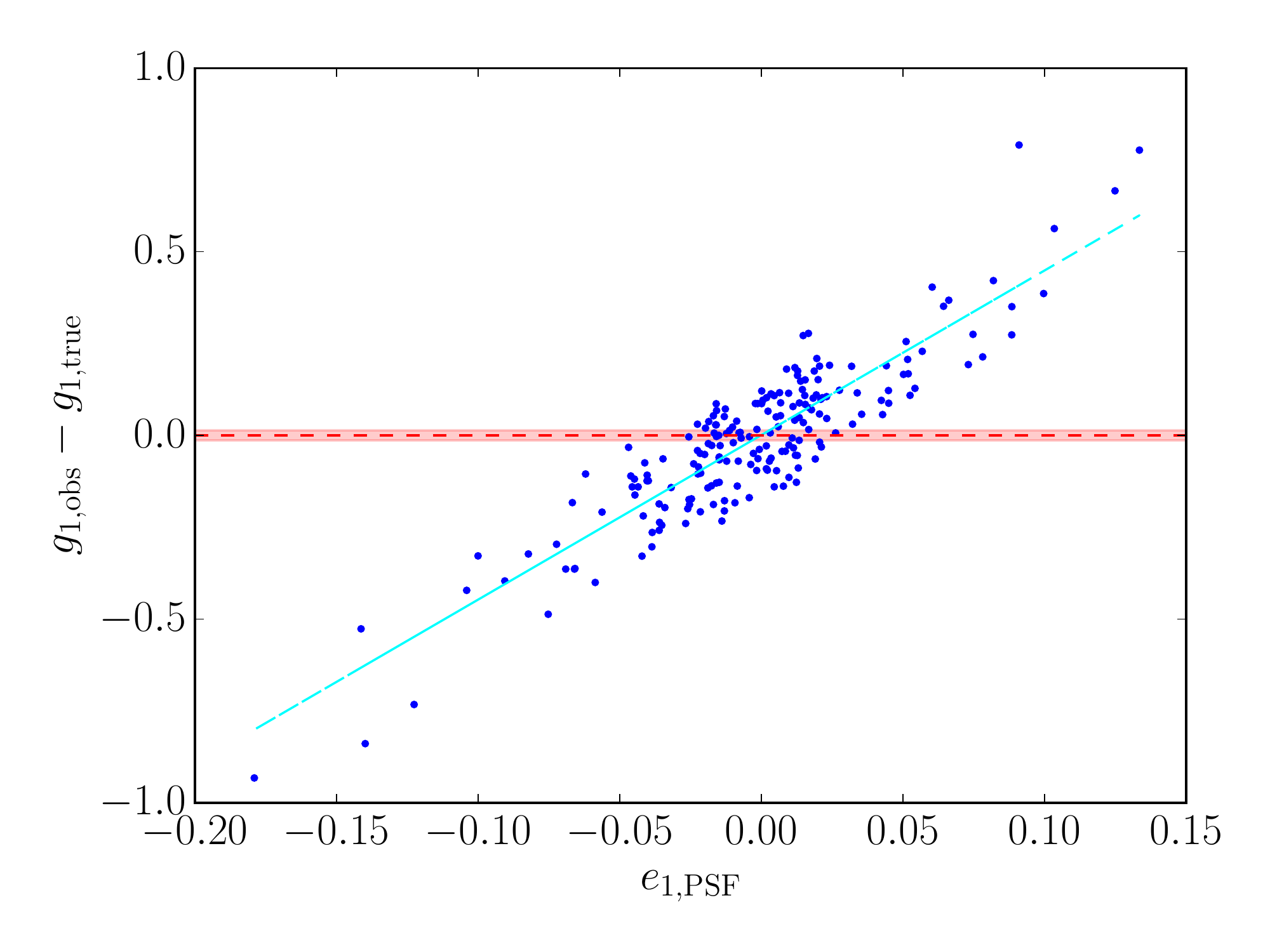}
\includegraphics[width=0.4\linewidth]{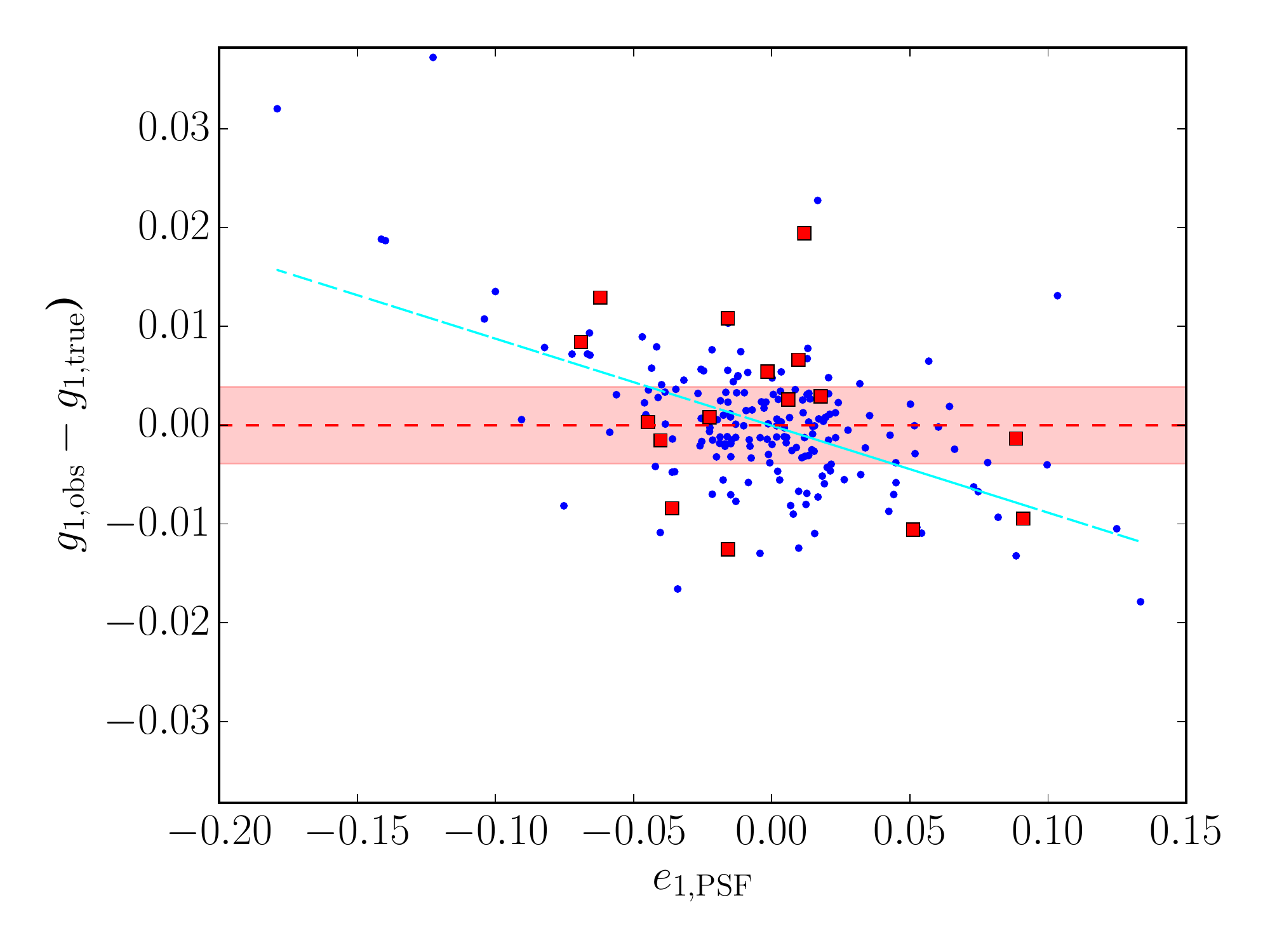}
\end{center}
\caption{Effects of the metacalibration algorithm applied to PSF
  correction. {\bf Left} panels show the relationship between measured
  shear and PSF ellipticity before correction, and {\bf right} panels
  show the same trends afterwards. Note that the shaded horizontal
  band covers the same vertical range in each panel. Points
  rejected by our likelihood cut are shown with red boxes.  \\
  Simulation branch/algorithm pairs shown in order from {\bf top} to
  {\bf bottom} are RGC-regauss, CGC-KSB, and CGC-moments. The
  combination of the metacalibration algorithm with our
  maximum-likelihood averaging procedure makes accurate corrections
  when the PSF ellipticities are small or comparable to the magnitude
  of the shear signal. It is clear that a large fraction of the trend
  remaining after correction is driven by remaining unmasked
  high-PSF-ellipticity outlier fields. While these were not rejected
  by our likelihood criterion, they would typically not pass the image
  quality requirements in a realistic experiment.}
\label{fig:psf_trends}
\end{figure*}

\section{Results}
Metacalibration reduces the shear calibration biases in every branch
that we have tested in, for all methods, and nearly in all cases to a level that is consistent with
zero within the errors. In the one branch (CGC-Moments) where it does not
eliminate multiplicative biases for both shear components within our
ability to measure them, it reduces $m$ by a factor of eighty. Our
objective criterion for the quality of per-field mean shear inference
generally succeeds in flagging problem fields, and (as is visible in
figure~\ref{fig:m_comparison}), the metacalibrated shears tend to be
less noisy than the un-metacalibrated shears.

Our PSF detrending algorithm is less successful, probably for reasons
we discuss in section~\ref{sec:appplicability}. Even here, most fields
see a significant reduction in the impact of the PSF anisotropy on the inferred
shears.

The remainder of section draws conclusions about the performance of the
metacalibration algorithm by comparing the results across simulation
branches. We discuss the impact of realistic galaxy morphologies,
correlated noise, a heterogeneous PSF, and shape measurement
algorithms.

The additive and multiplicative calibration biases for each simulation
branch, with and without metacalibration, are reported in full in
Table~\ref{table:results}. Before/after trends for the multiplicative
calibration biases and the trends with PSF ellipticity are shown in
figures~\ref{fig:m_comparison} and \ref{fig:psf_trends},
respectively. Visualizations of the overall impact of metacalibration
and PSF ellipticity detrending on the ensemble-average quantities in
each branch are provided in Figures~\ref{fig:m_results} and
\ref{fig:a_results}.


\subsection{Galaxy Morphology}
We can compare the performance of the metacalibration algorithm on
model and realistic galaxy morphologies in two cases (RGC-regauss vs.\
CGC-regauss, and RGC-noaber-regauss vs.\ CGC-noaber-regauss). In
neither case does the introduction of realistic morphology have any
impact on the multiplicative shear calibration biases: metacalibrating
both pairs of branches results in multiplicative biases that are
consistent with zero.  The most notable difference between the model
and real morphologies is apparent in the PSF trends. Our perturbative
PSF detrending scheme reduces the residual additive bias in most
cases, but it does not completely correct the model morphologies,
whereas there is no evidence of any residual PSF effects in any of the
realistic morphology branches.  This should not surprise us: we have
chosen to perturb and detrend the linear effects of PSF ellipticity,
but other PSF properties may be more significant and hence our model
is not a complete description of additive systematics. The coupling
between PSF morphology and inferred shear will in general depend on
the details of both, as well as the galaxy morphology and measurement
algorithm. We defer further development of methods to select
appropriate control variables for PSF detrending to future work.

\subsection{Correlated Noise}
The image reconvolution procedure described in
section~\ref{sec:counterfactual} also modifies the noise field in the
original image: the isotropic white noise typical of real images will
be transformed in the counterfactual images into a correlated noise
field with preferred direction and scale. 

This effect was first documented in the context of shear measurement
in the work described in \citet{metacalII} in simulations
conducted at much lower signal-to-noise ratios than those of the
Great3 challenge. A fuller exploration of the impact of correlated
noise on metacalibration, along with a comprehensive mitigation
strategy, is presented in that work.

We investigate the impact of this noise by successively adding noise
to the CGC images and re-running the metacalibration and shear
estimation algorithms. In the process we gradually increase the noise
relative to its initial level by up to a factor of $3.5$.

\begin{figure}
\includegraphics[width=\columnwidth]{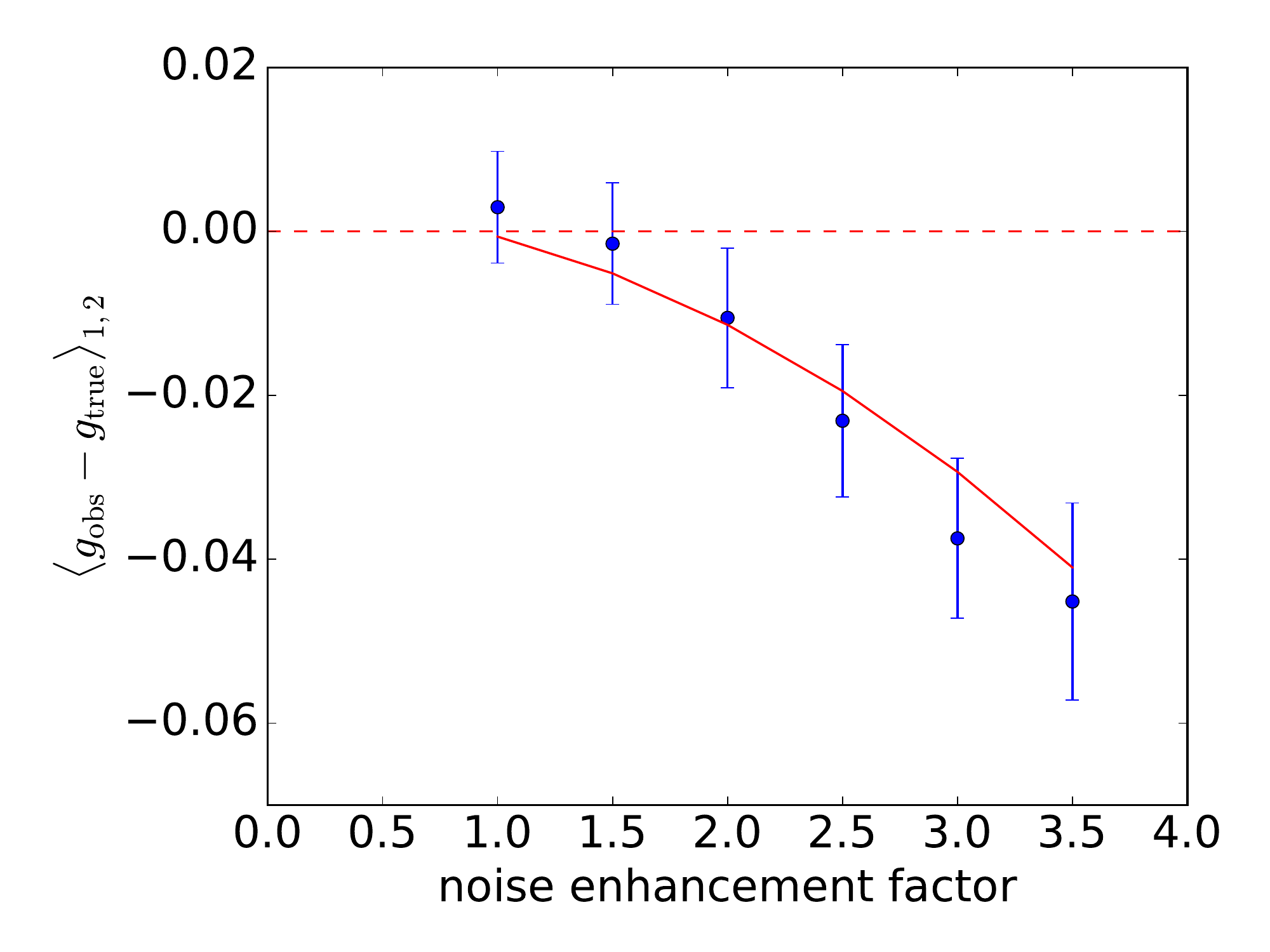}
\caption{Effects of introducting additional noise. Points with errors
  correspond to additive shear bias in the control-ground-constant
  branch when additional noise is added. The noise enhancement factor
  corresponds to the factor by which the noise in each galaxy image is
  increased relative to the fiducial GREAT3 simulations. The solid red
  line shows the expected power-law scaling resulting from correlated
  noise bias, with a normalization fixed to the measured additive
  biases.}
\label{fig:noise_bias}
\end{figure}

The additive biases resulting from varying the noise level are
shown in Figure~\ref{fig:noise_bias}. For the typical signal-to-noise ratios seen
in the GREAT3 simulations, biases resulting from correlated noise
appear to scale roughly as the variance of the initial noise field. We
note that these effects do not become significant until the overall
signal-to-noise ratio has been reduced by a factor of two compared to the fiducial
GREAT3 values.

\begin{figure*}[t]
\begin{center}
\includegraphics[width=0.8\linewidth]{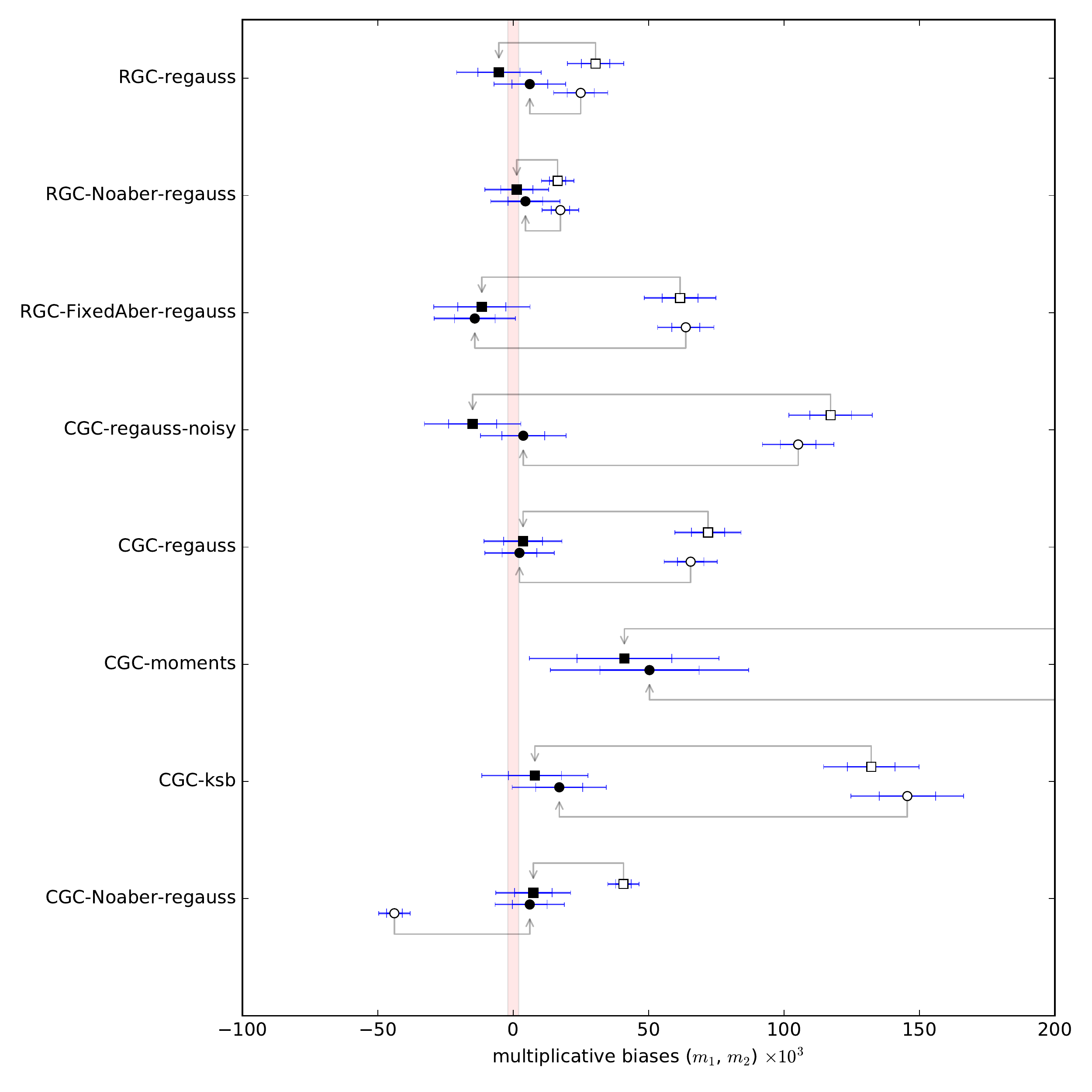}
\end{center}
\caption{Calibration bias results. Each row shows muliplicative
  calibration bias $m_1$ (top) and $m_2$ (bottom) before and after
  metacalibration. Pre- and post-correction points are connected by
  gray arrows: in every case the procedure has reduced or eliminated
  the amplitude of detectable multiplicative calibration bias.}
\label{fig:m_results}
\end{figure*}

\begin{figure*}[t]
\begin{center}
\includegraphics[width=0.8\linewidth]{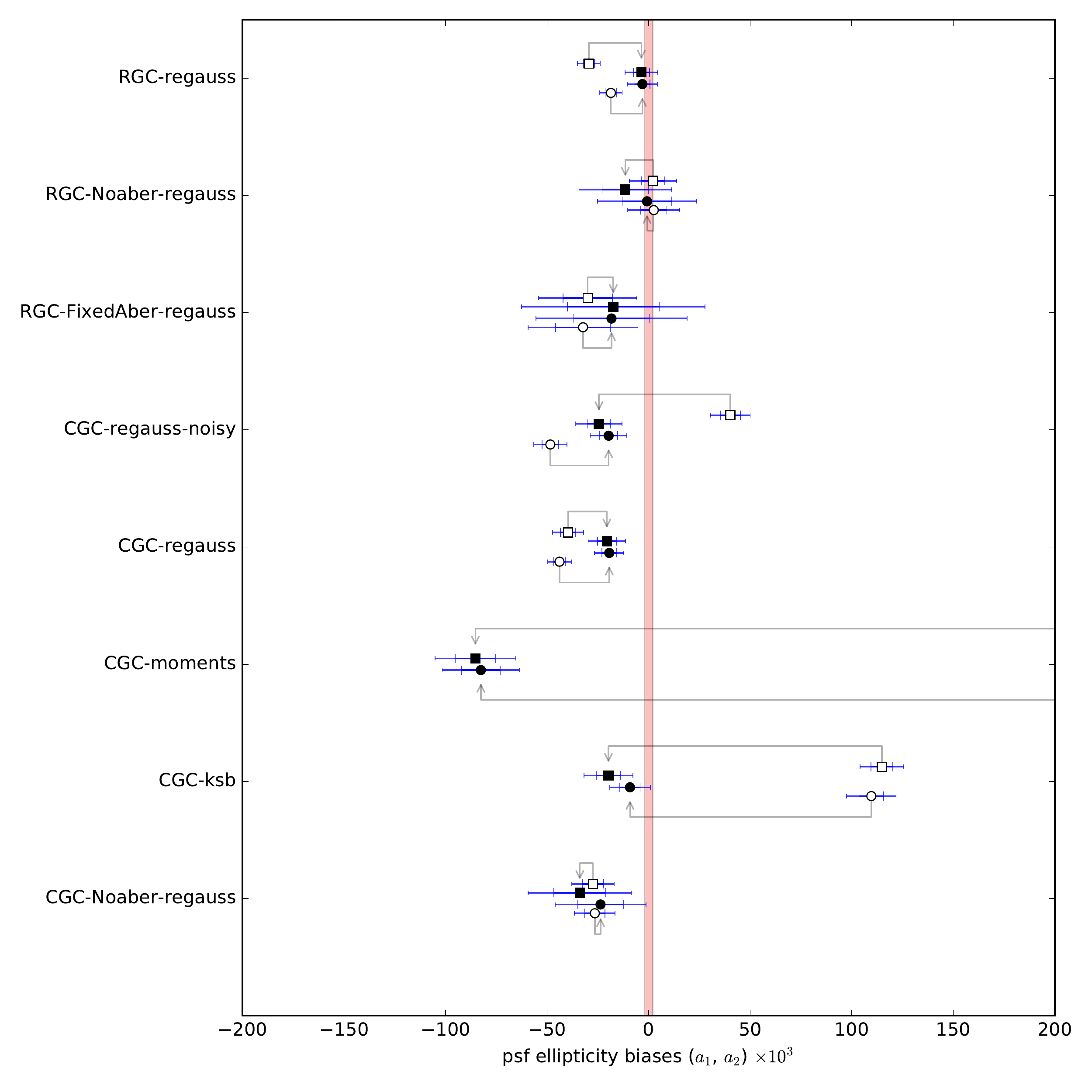}
\end{center}
\caption{Additive shear bias results. Each row shows linear coupling
  between the PSF ellipticity and the measured shape $a_1$ (top) and
  $a_2$ (bottom) before and after metacalibration. Pre- and
  post-correction points are connected by gray arrows: in every case
  the procedure has reduced or eliminated the amplitude of detectable
  PSF coupling.}
\label{fig:a_results}
\end{figure*}

The case of the factor of two increase is denoted ``CGC-regauss-noisy'' in
Table~\ref{table:results} and Figures~\ref{fig:m_results} and
\ref{fig:a_results}.

\subsection{Heterogeneous PSF}
Large variations in the PSF properties can impact our measurement
algorithm via two channels: first, a heterogeneous PSF can result in
individual fields deviating from the zero-shear distribution
constructed from the ensemble of measured shapes, potentially biasing
the histogram estimator; and second (as discussed above) via the usual
mechanism of incomplete PSF correction and detrending.

Eliminating outlier fields in those branches with large variations in
the PSF via the rejection mechanism described in
section~\ref{sec:model_checking} tends to significantly improve the
calibration bias after metacalibration. The rejected fields (red
squares in Figure~\ref{fig:psf_trends}) tend to have substantially
larger residual shear calibration biases than the mean field in each
branch, and without the outlier rejection, we see few-percent level
calibration biases in each of these branches. With the likelihood-based 
rejection mechanism in place, the multiplicative and additive biases
in the CGC and RGC-regauss branches are consistent with those in the
CGC-noaber and RGC-noaber branches. It should be noted that we arrived
at the $10\%$ cut by choosing a level that typically eliminated
outliers from the compact core of the distribution of per-field
likelihoods. There are certainly more objective ways to choose this
cut -- one can eliminate fields that are much lower than the values
generated by sampling from equation~\eqref{eqn:multnomial}, for instance
-- but we defer calibration of the outlier-rejection technique to future work.

\begin{figure}
\begin{center}
\includegraphics[width=0.8\columnwidth]{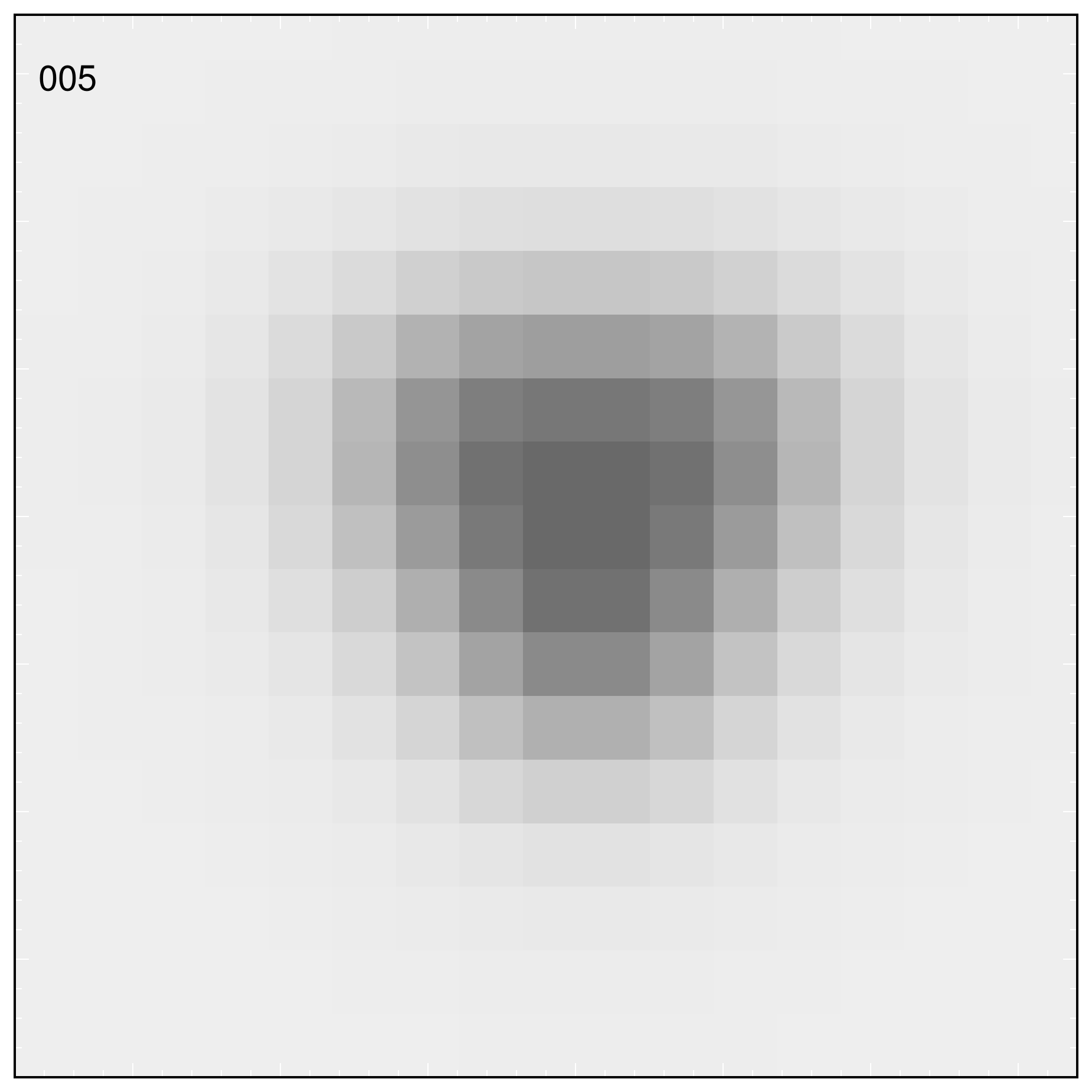}
\end{center}
\caption{Example of a PSF in one of the simulations with fixed, large aberrations.  The most obvious
feature in this case is the trefoil, which gives rise to the triangular shape. \label{fig:trefoil}}
\end{figure}

Finally, we compare the performance of metacalibration in the
RGC-noaber, and RGC-FixedAber branches. These branches have a
relatively homogenous PSF, differing only in the complexity of the PSF
morphology (see Figure ~\ref{fig:trefoil}). We see only marginal
evidence ($\sim1.5\sigma$) for residual calibration biases in this
case (though this is the only significant bias seen among the six
regauss fields we report on, so a single $\sim1\sigma$ detection
is to be expected even in the absence of any real calibration
biases). It is also notable that the RGC-FixedAber branch sees no
significant trend between PSF anisotropy and inferred shear.

Given these results, it would seem that the primary channel through
which a heterogeneous PSF impacts our shear inference is via the first
channel; the ensemble response of those fields with outlier PSF
properties is sufficiently different from that estimated from the
global null ellipticity distribution that identification of these fields
is essential.

\subsection{Shape Measurement Algorithms}
We investigate the effects of the shape measurement algorithm by
comparing the performance of the regauss, KSB, and linear moments
measurement algorithms on the CGC simulation branch. We see large
improvements in performance for each method after using
metacalibration. There is weak evidence for residual KSB calibration
biases, and strong evidence for residual biases in the linear
moments. In the latter case, however, the improvement relative to the
un-metacalibrated case is large: the multiplicative and additive
biases are reduced by factors of $80$ and $51$, respectively,
resulting in performance for metacalibrated linear moments that is
quantitatively superior to that of the un-metacalibrated KSB.  This is
remarkable given that simple linear moments involve no PSF correction,
leaving the entirety of the PSF correction process to metacalibration.

One potential reason for the difference in performance for different
shape measurement methods is the linearity of the ensemble estimates
in the resulting shear. We find values for $\eta_{nl}$ of
$-0.0034\pm0.0004$, $-0.024\pm0.002$, and $-0.0037\pm0.0004$, for the
regauss, KSB, and linear moments algorithms, respectively (where the
errors are bootstrap estimates). The quadratic correction to the shear
response is much larger for KSB than for the other methods, suggesting
that any problems with the KSB shapes may arise from its more
nonlinear shear response (see \citealt{2011MNRAS.410.2156V} for a
detailed investigation of this issue).


Another potential driver of residual additive systematics is
incomplete PSF detrending. The reasons for this are discussed in the
previous section. The linear moments see significant PSF residuals,
and Figure~\ref{fig:psf_trends} suggests that these are primarily
driven by the fields with large PSF ellipticities.

The $\eta_{nl}$ values and the residual additive and multiplicative
biases for these methods suggest that the small residual calibration
bias in KSB is driven by the nonlinear response of that shape
measurement algorithm, while the residual linear moments calibration
bias is primarily driven by imperfect PSF correction.

\begin{table*}
\centering
\makebox[0.8\textwidth][c]{
\begin{tabular}{l  c |cc | cc |cc }
\hline
branch & algorithm  &  $m_1\times1000$ & $m_2\times1000$  & $a_1\times1000$ & $a_2\times1000$ &  $c_1\times1000$ & $c_2\times1000$ \\
\hline
\hline
CGC & regauss (MC)  & $3.6 \pm 7.2$ & $2.3\pm 6.4$ & $-20.6\pm4.6$ & $-19.4\pm3.6$ &  $0.0\pm.2$ & $0.1\pm0.2$ \\
CGC & regauss  &  $71.9\pm6.1$ & $65.5\pm4.9$  & $-39.7\pm3.8$ & $-43.9\pm2.9$ &  $0.0\pm0.1$ & $0.0\pm0.1$ \\
CGC & regauss-noisy (MC) & $-15.0\pm 8.9$ & $3.7\pm7.9$ & $-24.5\pm5.7$ & $-19.7\pm4.5$ & $-0.1\pm0.2$ & $-0.1\pm0.2$\\
CGC & regauss-noisy & $117.2 \pm  7.7$ & $105.2\pm6.6 $ & $40.2\pm4.9$ & $-48.4\pm4.1$ & $0.2\pm0.2$ & $-0.1\pm0.2$ \\
CGC & ksb (MC)  &  $8.0\pm9.8$ & $-17.0\pm8.7$  & $-19.8\pm6.0$ & $-9.2\pm5.0$ &  $0.3\pm0.2$ & $-0.2\pm0.2$ \\
CGC & ksb  &  $132.2\pm8.8$ & $145.5\pm10.4$  & $114.8\pm5.4$ & $109.6\pm6.1$ &  $-0.4\pm0.2$ & $0.0\pm.3$ \\
CGC & moments (MC) & $37.1\pm17.3$ & $45.9\pm18.4$ & $-88.0\pm9.9$ & $-87.7\pm9.5$ & $0.1\pm0.4$ & $-0.2\pm0.5$\\
CGC & moments & $3417.0\pm138.2$ & $3223.8\pm173.1$ & $4480.5\pm77.6$ & $4604.4\pm88.8$ & $0.9\pm3.4$ & $-0.8\pm4.6$ \\
RGC & regauss (MC) &  $-5.3\pm7.8$ & $6.1\pm6.6$  & $3.6\pm4.0$ & $-3.1\pm3.7$ &  $0.1\pm0.2$ & $0.1\pm0.2$ \\
RGC & regauss &  $30.4\pm5.2$ & $24.9\pm5.0$  & $-29.5\pm2.8$ & $-18.6\pm2.8$ &  $0.0\pm0.1$ & $0.2\pm0.1$ \\
CGC-Noaber & regauss (MC)  &  $7.4\pm6.9$ & $6.1\pm6.9$  & $-33.9\pm12.7$ & $-23.7\pm11.2$ &  $-0.1\pm0.2$ & $0.1\pm0.2$ \\
CGC-Noaber & regauss  &  $40.7\pm2.9$ & $43.9\pm2.9$  & $-27.4\pm5.2$ & $-26.5\pm5.0$ &  $0.1\pm0.1$ & $-0.1\pm0.1$ \\
RGC-Noaber & regauss (MC)&  $-1.3\pm5.9$ & $4.5\pm6.4$  & $-11.5\pm11.4$ & $-0.8\pm12.2$ &  $0.0\pm0.2$ & $-0.1\pm0.2$ \\
RGC-Noaber & regauss &  $16.4\pm3.0$ & $17.4\pm3.4$  & $2.2\pm5.8$ & $2.5\pm6.4$ &  $0.2\pm0.1$ & $0.0\pm0.1$ \\
RGC-FixedAber & regauss (MC) &  $-11.6\pm8.9$ & $-14.2\pm7.5$  & $17.4\pm22.6$ & $18.3\pm18.6$ &  $0.2\pm0.2$ & $-0.2\pm0.2$ \\
RGC-FixedAber  & regauss  &  $61.6\pm6.6$ & $63.7\pm5.2$  & $-30.0\pm12.1$ & $-32.3\pm13.5$ &  $0.3\pm0.2$ & $0.0\pm0.1$ \\
\hline
\end{tabular}
}
\caption{Shear and PSF calibration bias parameters from each of the
  branches considered. Rows with metacalibrated parameters are shown
  above their un-calibrated counterparts. }
\label{table:results}
\end{table*}

\section{Applicability to real data}
\label{sec:appplicability}
Several implementation issues need to be solved before this method can
be deployed on real survey data. We have made no attempt to deal with
the effects of masked pixels or blending, and while it seems clear
that our proposed algorithm has the potential to deal with selection
biases, we have not demonstrated that capability here. We have also
demonstrated the presence of a calibration bias which scales with the
variance of the pixel noise, which may be a result of the correlations
imposed on the noise field during the construction of the
counterfactual image.

In this work, we have attempted to remove PSF systematics by measuring
the response of the shape measure to PSF ellipticity. There is no
guarantee, however, that the PSF ellipticity is the correct parameter
to use for this detrending. In a realistic measurement, it would be
best to first determine which modes of PSF variation are most likely
to impact the chosen shape measure, and then use the image
modification and detrending technique described here to remove those
effects in the data.

Finally, our shear inference procedure is designed to extract the mean
shear from a constant-shear field. While this procedure may be
applicable to galaxy-galaxy lensing (i.e., projecting the shapes onto
the tangent to the appropriate lens), it is not suitable for
measurements like cosmic shear that typically rely on second or higher
moments of the shear field. While a similar histogram estimation
procedure could be implemented to model the responsivity of the
distribution of ellipticity products (as would be needed for two-point
shear correlation functions), we leave design and implementation of
this generalization to future work.

At the time of this writing, metacalibration is being actively adapted
to realistic measurements in the Dark Energy Survey. Concurrent work
\citep{metacalII} will demonstrate algorithmic improvements that
allow this technique to be used on Dark Energy Survey data with
state-of-the-art shear estimation algorithms.

\section{Conclusions}
We have proposed and implemented the first method for self-calibration
of shear measurements that does not rely on deeper calibration fields or
simulations. Our method can be wrapped around
any sufficiently well-behaved shape measurement algorithm.  We use
GREAT3 and related simulations to demonstrate that metacalibration
reduces or eliminates shear calibration biases across a variety of
galaxy morphologies, PSF properties, and for several otherwise biased
shape measurement algorithms. We have argued that our technique works
because it takes advantage of the fundamental linearity of
astronomical images in the weak lensing shear signal, in
combination with the fact that the effects of shear on an image with a
known PSF are model-independent.

Those cases we have examined where the initial biases are large or not
linear are not completely corrected by our linear detrending scheme,
though in every case we have studied the algorithm appears to
substantially improve biases resulting from faulty additive PSF correction and
multiplicative shear calibration biases. Even the nearly information-free linear moments
algorithm appears to be calibrated by our procedure to a level
superior to uncorrected KSB, a widely used traditional shear
estimation algorithm.

We expect future work to extend this method to deal with the
complexities inherent in real data.

\section*{Acknowledgments}

We are grateful to Erin Sheldon and to Mike Jarvis for
discussion, advice, and for checking the results reported here against
their own independent image simulation work.

We also thank Klaus Honscheid and Peter Melchior for useful
discussions and advice, and Neha Bagchi for proofreading and editing
support.

RM acknowledges the support of the Department of Energy Early Career
Award program.

EMH was supported for part of the work on this proposal by the US
Department of Energy’s Office of High Energy Physics
(DE-AC02-05CH11231), and from the Center for Cosmology and
AstroParticle Physics at the Ohio State University.

Part of the research was carried out at the Jet Propulsion Laboratory
(JPL), California Institute of Technology, under a contract with the
National Aeronautics and Space Administration.

\bibliographystyle{apj}
\bibliography{bibliography}

\end{document}